\documentclass[aps,prb,reprint,twocolumn,superscriptaddress,showpacs,floatfix]{revtex4-2}
\usepackage{amsmath,graphicx}
\usepackage[utf8]{inputenc}
\usepackage[T1]{fontenc}
\usepackage{xcolor}
\usepackage{textcomp}
\usepackage{bm}

\usepackage{physics}
\usepackage{amsmath}
\usepackage{tikz}
\usepackage{mathdots}
\usepackage{yhmath}
\usepackage{cancel}
\usepackage{color}
\usepackage{array}
\usepackage{multirow}
\usepackage{amssymb}
\usepackage{gensymb}
\usepackage{tabularx}
\usepackage{extarrows}
\usepackage{booktabs}
\usetikzlibrary{fadings}
\usetikzlibrary{patterns}
\usetikzlibrary{shadows.blur}
\usetikzlibrary{shapes}

\usepackage{mathtools}

\usepackage{color}
\definecolor{LinkColor}{rgb}{0.75,0.0,0.2}

\usepackage{hyperref}
\hypersetup{
	pdfauthor={good guys},
	pdftitle={good title},
	colorlinks=true,
	citecolor=LinkColor,
	linkcolor=LinkColor,
	urlcolor=LinkColor,
}

\usepackage{listings}
\definecolor{lightgray}{gray}{1}

\usepackage{theorem}

\usepackage{tikz}

\newcommand{\nc}{\newcommand}
\nc{\braoprket}[3]{\langle#1|#2|#3\rangle}
\nc{\opn}[1]{\operatorname{#1}}
\nc{\avg}[1]{\langle#1\rangle}
\nc{\ketbrasame}[1]{|#1\rangle\!\langle#1|}
\nc{\swap}{\opn{SWAP}}
\nc{\E}{\mathbb{E}}
\nc{\Var}{\opn{Var}}
\nc{\dg}{\dagger}

\usepackage[normalem]{ulem}

\begin{document}

\title{Topological tricritical Ising universality class in one dimension}

\author{Sheng Yang}
\affiliation{Institute for Advanced Study in Physics and School of Physics, Zhejiang University, Hangzhou 310058, China}


\author{Hai-Qing Lin}
\affiliation{Institute for Advanced Study in Physics and School of Physics, Zhejiang University, Hangzhou 310058, China}

\author{Xue-Jia Yu}
\email{xuejiayu@eitech.edu.cn}
\affiliation{Eastern Institute of Technology, Ningbo 315200, China}

\begin{abstract}

Quantum critical points can host symmetry-protected topological edge modes even when the bulk is gapless, giving rise to symmetry-enriched universality classes beyond the conventional Landau--Ginzburg--Wilson paradigm.
Here we show that the tricritical Ising (TCI) critical point---described by a paradigmatic conformal field theory (CFT) with emergent supersymmetry---admits a topologically nontrivial, symmetry-enriched realization, which we term the topological TCI.
We construct this critical point in the cluster O'Brien--Fendley spin chain by applying a symmetry-protected topological entangler to the original O'Brien--Fendley model.
The two realizations therefore share the same bulk TCI CFT.
Under open boundary conditions, however, the original model exhibits the free conformal boundary condition, whereas the cluster model exhibits a spontaneously fixed boundary condition: an infinitesimal boundary field can select one of the fixed boundary states, yielding spontaneous boundary magnetization.
We further show that symmetry enrichment and boundary renormalization-group flow provide two physically distinct routes to the spontaneously fixed boundary condition.
Whereas the free and spontaneously fixed boundary conditions of the original model can be interconverted by tuning a symmetry-preserving boundary term, no such conversion occurs for the cluster model along the boundary deformations studied here, even in the strong-deformation limit.
Although they share the same conventional boundary CFT data, including the boundary operator content and boundary $g$-function, the two realizations of the spontaneously fixed boundary condition are distinguished by different time-reversal charges of the disorder field.
For the tested symmetry-preserving boundary perturbations that lift the exact twofold degeneracy, the finite-size energy splitting is numerically consistent with exponential decay over accessible system sizes, in stark contrast to the algebraic splitting at the symmetry-enriched Ising critical point of the cluster Ising spin chain.
We also find that the topological TCI realization persists along a critical line in a two-parameter phase diagram.
These results establish a symmetry-enriched TCI universality class and uncover boundary distinctions invisible to conventional boundary CFT data.

\end{abstract}

\maketitle

\section{Introduction}

Universality is a central organizing principle of modern theoretical physics.
Within the Landau--Ginzburg--Wilson paradigm, continuous phase transitions are organized into universality classes according to dimensionality, symmetry-breaking patterns, and order-parameter structure, with each class characterized by universal critical exponents~\cite{landau2013statistical,Sachdev_1999,Sondhi1997RMP}.
Recent developments~\cite{YU20261}, however, have shown that this characterization can be incomplete for quantum critical points (QCPs).
In the presence of symmetry, the same underlying critical behavior can in principle admit distinct symmetry-enriched realizations, so that additional symmetry-resolved information may be required to fully characterize a QCP.
This question is not restricted to conformal critical points~\cite{Duque2021PRB,yang2026topologicalquantumcriticalityquasiperiodic};
nevertheless, conformal criticality provides a particularly controlled setting in which such additional universal information can be identified and characterized explicitly.

In a gapped symmetry-protected topological (SPT) phase~\cite{gu2009prb,pollmann2012prb}, the nontrivial symmetry structure of a short-range-entangled bulk gives rise to boundary degrees of freedom that cannot be removed without breaking the protecting symmetry or closing the bulk gap.
Remarkably, related edge structures can survive when the bulk becomes gapless, leading to topologically nontrivial, or symmetry-enriched, QCPs~\cite{Verresen2018PRL,verresen2020topologyedgestatessurvive,Verresen2021PRX,Duque2021PRB,Yu2022PRL,Mondal2023PRB,Yu2024PRB,Prembabu2024PRB,Zhong2025PRB,Huang2025SciPost,Li2025PRB,Rey2025PRB,Zhou2025CP,chou2025ptsymmetryenrichednonunitarycriticality,guo2026generalizedlihaldanecorrespondencecritical,deng2026anomalousdynamicalscalingtopological,prembabu2025noninvertibleinterfacessymmetryenrichedcritical,yang2026topologicalquantumcriticalityquasiperiodic,TanCP2026,chen2026criticaltopologicalphotonicssynthetic} and, more broadly, gapless SPT states~\cite{Keselman2015PRB,Scaffidi2017PRX,JIANG2018753,Parker2018PRB,Thorngren2021PRB,Hidaka2022PRB,Wen2023PRB,Yu2024PRL,Zhang2024PRA,Su2024PRB,Li2025SciPost,Wen2025PRB,Yang2025CP,Flores2025PRL,yang2025deconfinedcriticalityintrinsicallygapless,prembabu2025multicriticalitypurelygaplessspt,Yu2026PRB,xu2026frameworkpredictingentanglementspectra,chou2026topologicallyenforcedlifshitzmulticriticality}.
For QCPs described by CFTs~\cite{Verresen2018PRL,Verresen2021PRX,Yu2022PRL}, the symmetry properties of the low-energy CFT can provide additional discrete topological invariants.
In particular, the symmetry charges carried by symmetry fluxes, represented microscopically by nonlocal string operators, can distinguish different symmetry-enriched realizations of the same CFT and imply distinct boundary phenomena.
These distinctions can produce different conformal boundary conditions~\cite{Yu2022PRL,Parker2018PRB}, algebraically localized edge modes~\cite{Verresen2021PRX,Yang2025CP}, and nontrivial bulk-boundary correspondences~\cite{Yu2024PRL,Zhong2025PRB,guo2026generalizedlihaldanecorrespondencecritical}.

A basic example is the topologically nontrivial Ising, or Ising*, critical point of the cluster Ising chain~\cite{Verresen2021PRX,Yu2022PRL}.
A finite-depth SPT entangler relates the cluster Ising chain to the standard transverse-field Ising chain under periodic boundary conditions (PBC), so they realize the same bulk Ising CFT.
Nevertheless, their nonlocal disorder operators carry different charges under spinless time reversal, and their natural open boundaries exhibit different edge structures~\cite{Verresen2021PRX}.
Here and throughout, the asterisk labels the symmetry-enriched realization rather than a different bulk CFT.
Moreover, symmetry-preserving boundary perturbations can couple the symmetry-protected edge degrees of freedom through the critical bulk, producing an algebraic finite-size energy splitting in the Ising* case~\cite{Verresen2021PRX}.
This example naturally raises the question of how the framework of symmetry-enriched criticality extends beyond the Ising CFT and what new physics may thereby be uncovered.
In particular, it is important to understand which features of the Ising$^*$ critical-edge physics are universal and which depend on the operator structure of the underlying critical theory.
The tricritical Ising (TCI) universality provides a particularly natural non-Ising setting for addressing this question. 
It occurs where a line of continuous Ising transitions meets a first-order transition, and is described by the unitary Virasoro minimal model $\mathcal{M}(4,5)$ with central charge $c=7/10$~\cite{francesco2012conformal}.
In addition to possessing a richer spectrum of bulk and boundary operators than the Ising CFT, the TCI CFT exhibits emergent supersymmetry~\cite{Friedan1984PRL,FRIEDAN198537,FODA1988611,Fendley2003PRL,Lee2007PRB,Bauer2013PRB,Grover2014Science,Huijse2015PRL,Rahmani2015PRB,Rahmani2015PRL,Jian2015PRL,Ejima2016PRB,Zhu2016PRB,Li2017PRL,Jian2017PRL,Li2018SA,OF2018PRL,Ebisu2019PRL,Yu2022PRB,Roychowdhury2024PRR,Miura2024PRB,Li2024PRL,Liu2025PRB,Miura2025PRB,samanta2025realizingsupersymmetrydigitizedquantum,wu2025emergentspacetimesupersymmetry2d,sun2026experimentalobservationconformalfield}.
These features raise the central question of whether topological TCI criticality can exist and, if so, whether its boundary physics behaves differently from that of Ising*.

To explore this possibility, we consider the O'Brien--Fendley (OF) spin-$1/2$ chain, whose Ising critical line terminates at a TCI point beyond which the transition becomes first order~\cite{OF2018PRL}.
The cluster OF model is obtained from the original OF model by applying the same type of finite-depth SPT entangler used in the cluster-Ising construction.
Both the cluster and original OF models preserve a global $\mathbb{Z}_2$ spin-flip symmetry and a spinless antiunitary time-reversal symmetry $\mathbb{Z}_2^T$, which together form the protecting symmetry group $\mathbb{Z}_2 \times \mathbb{Z}_2^T$.
Under PBC, the two Hamiltonians are exactly related by conjugation with the SPT entangler.
Consequently, their local bulk CFT data at tricriticality, including the central charge and local bulk critical exponents, are exactly identical.
Under open boundary conditions (OBC), however, this equivalence does not extend to the natural boundary terms, and the cluster OF model hosts symmetry-protected topological edge modes that give rise to a twofold boundary degeneracy.
We refer to this symmetry-enriched realization of the TCI CFT as topological TCI, or TCI*.
The distinction between the two realizations therefore appears in the symmetry properties of nonlocal disorder operators and in boundary physics.

Boundary conformal field theory (BCFT) provides the natural framework for characterizing boundary critical behavior.
Even when two systems share the same bulk CFT, different conformal boundary conditions can lead to distinct boundary operator content and boundary responses~\cite{francesco2012conformal,cardy2008boundaryconformalfieldtheory}.
Both the Ising and TCI BCFTs admit free, fixed, and spontaneously fixed boundary conditions, which can be qualitatively distinguished by how the global $\mathbb{Z}_2$ symmetry acts on them.
At a free boundary, the boundary condition preserves the global $\mathbb{Z}_2$ symmetry, and the boundary magnetization vanishes.
At a fixed boundary, the $\mathbb{Z}_2$ symmetry is explicitly broken by fixing the boundary spin in one of the two symmetry-related directions.
At a spontaneously fixed boundary, by contrast, the boundary Hamiltonian preserves the $\mathbb{Z}_2$ symmetry, but the boundary spontaneously breaks it in the thermodynamic limit, giving rise to two degenerate boundary states with opposite nonzero magnetizations that are exchanged by the $\mathbb{Z}_2$ symmetry;
an infinitesimal symmetry-breaking boundary field would select one of these two states~\cite{Ian_Affleck_2000,Iino2020PRB,Verresen2021PRX}.
The original OF model naturally realizes the free boundary condition, but can also realize the spontaneously fixed boundary condition when the strength of a symmetry-preserving boundary term is tuned across a boundary phase transition controlled by an unstable degenerate boundary condition~\cite{Iino2020PRB}.
This comparison first raises the question of whether the natural open-boundary termination of the cluster OF chain realizes the spontaneously fixed boundary condition, as in the Ising* case~\cite{Verresen2021PRX}.
If so, three further questions follow: whether the natural cluster realization is physically distinct from the boundary-tuned realization in the original OF model despite their identical conventional BCFT data;
whether the associated boundary degeneracy persists under finite and strong symmetry-preserving boundary perturbations;
and how the resulting finite-size energy splitting compares with the algebraic splitting at the Ising* critical point.

In this work, we answer these questions using a combination of density-matrix renormalization-group (DMRG) numerical simulations~\cite{White1992PRL,SCHOLLWOCK201196} and BCFT analysis.
From the boundary magnetization, boundary-bulk correlations, and low-energy open-chain spectroscopy, we show that the natural open-boundary termination of the cluster OF chain realizes a spontaneously fixed boundary condition that hosts a twofold topological degeneracy protected by the $\mathbb{Z}_2 \times \mathbb{Z}_2^T$ symmetry, in contrast to the free boundary condition realized by the original OF model.
We further demonstrate that the spontaneously fixed boundary condition has two physically distinct realizations in the TCI CFT: one arises from symmetry enrichment in the cluster OF model, whereas the other is reached through boundary renormalization-group (RG) flow in the original OF model.
Although they share the same conventional BCFT data, including the boundary operator content and boundary $g$-function~\cite{affleck1991prl}, they are distinguished by the $\mathbb{Z}_2^T$ charge of the disorder field.
For the cluster OF model, we also show that the boundary degeneracy persists as the strength of the symmetry-preserving boundary terms studied here is varied from weak to strong.
For the cases in which these terms lift the exact twofold degeneracy, the finite-size energy splitting is numerically consistent with exponential decay over accessible system sizes, in stark contrast to the algebraic splitting at the Ising* critical point~\cite{Verresen2021PRX}.
Finally, we find that topological TCI persists along a  critical line in a two-parameter phase diagram.
Our results establish a symmetry-enriched TCI universality class and uncover boundary distinctions invisible to conventional BCFT data.

The remainder of the paper is organized as follows.
Section~\ref{sec:model-bcft} introduces the original and cluster OF models and reviews the TCI CFT and BCFT background.
Section~\ref{sec:natural-boundary} identifies the natural boundary condition of the cluster OF model, while Sec.~\ref{sec:finite-size-splitting} examines the finite-size energy splitting of its boundary degeneracy under symmetry-preserving boundary perturbations.
Section~\ref{sec:symmetry-enriched-boundaries} compares the symmetry-enriched and boundary-RG realizations of the TCI spontaneously fixed boundary condition.
Section~\ref{sec:extended-tci-line} constructs the extended TCI* critical line, and Sec.~\ref{sec:conclusion} summarizes our main findings.
Appendices~\ref{app:numerical-methodology}-\ref{app:obc-susy} provide details of the DMRG simulations, complementary TCI and Ising CFT analyses, numerical checks of the bulk criticality and supersymmetry, the edge-mode splitting and boundary RG analyses, additional results for the $ZZ$-type boundary deformation, and the open-chain supersymmetry analysis.

\section{Cluster OF Model and TCI BCFT}
\label{sec:model-bcft}

\subsection{Original OF model and TCI CFT}

We begin with the original OF spin-$1/2$ chain proposed in Ref.~\cite{OF2018PRL};
its Hamiltonian reads
\begin{equation}
  \begin{aligned}
    H_{\rm OF}^{} = {} & - \sum_i \left( X_i + Z_iZ_{i+1} \right) \\
                       & + g \sum_i \left( X_iZ_{i+1}Z_{i+2} + Z_iZ_{i+1}X_{i+2} \right) \, .
  \end{aligned}
  \label{eq:of-model}
\end{equation}
Here $X_i$ and $Z_i$ are Pauli matrices, and $g$ is the coupling strength that controls the three-spin interactions;
it should not be confused with the Affleck--Ludwig boundary $g$-function calculated below.
At $g = 0$, the model reduces to the critical transverse-field Ising chain, for which Kramers--Wannier duality interchanges the transverse-field and Ising-exchange terms.
The $g$-dependent interactions preserve the self-duality precisely and tune the model along the continuous transition line toward its tricritical endpoint.

Along this self-dual line, the original OF model is described by the Ising CFT for $g < g_c$ and reaches a tricritical endpoint near $g_c \simeq 0.428$, beyond which the transition becomes first order~\cite{OF2018PRL}.
The endpoint is described by the TCI CFT, the unitary minimal model $\mathcal{M}(4,5)$ with central charge $c=7/10$~\cite{francesco2012conformal}.
Its six diagonal Virasoro primary fields and their conformal weights and scaling dimensions are summarized in Table~\ref{tab:tci-bulk-primaries}.
For these diagonal bulk fields, $h_\phi = \bar h_\phi$ and $\Delta_\phi = 2h_\phi$.
Further details of the Virasoro algebra and emergent superconformal structures of the TCI CFT are provided in Appendix~\ref{app:bcft}.

\begin{table}[t]
  \centering
  \caption{Diagonal bulk primary fields of the TCI CFT.}
  \label{tab:tci-bulk-primaries}
  \renewcommand{\arraystretch}{1.15}
  \begin{tabular}{c@{\qquad}c@{\qquad}c}
    \hline\hline
    Field $\phi$ & $h_\phi = \bar h_\phi$ & $\Delta_\phi$ \\
    \hline
    $I$          & $0$    & $0$    \\
    $\epsilon$   & $1/10$ & $1/5$  \\
    $\epsilon'$  & $3/5$  & $6/5$  \\
    $\epsilon''$ & $3/2$  & $3$    \\
    $\sigma$     & $3/80$ & $3/40$ \\
    $\sigma'$    & $7/16$ & $7/8$  \\
    \hline\hline
  \end{tabular}
\end{table}

The fields used directly in this work are the spin field $\sigma$, the energy field $\epsilon$, the fermionic fields associated with the emergent supersymmetry, and the disorder field $\mu$.
The leading lattice representative of the spin field is $\sigma_i \propto Z_i$.
The original-OF lattice representatives used for the bosonic energy field and the fermionic field are~\cite{zou2020prb}
\begin{equation}
  \begin{split}
    \epsilon_i & \propto \frac{1}{2} ( Z_{i-1}Z_i + Z_iZ_{i+1} ) - X_i \, , \\
    \psi_i & \propto \left[ \prod_{m=-\infty}^{i-1} X_m \right] ( Z_i + Y_i ) \, .
  \end{split}
  \label{eq:of-lattice-fields}
\end{equation}
The energy field has $\Delta_\epsilon = 1/5$.
Its fermionic partners $\psi$ and $\bar\psi$ have conformal weights $(h, \bar h)_{\psi} = (3/5, 1/10)$ and $(h, \bar h)_{\bar\psi} = (1/10, 3/5)$, respectively, and hence scaling dimension $\Delta_\psi = \Delta_{\bar\psi} = 1/10 + 3/5 = 7/10$~\cite{francesco2012conformal,zou2020prb,Friedan1984PRL,FRIEDAN198537,FODA1988611}.
They are fermionic components of the TCI supermultiplet rather than additional fields in the diagonal-primary list of Table~\ref{tab:tci-bulk-primaries}.
The nonlocal disorder field has the same scaling dimension as the spin field, $\Delta_\mu = 3/40$, and in the original OF model is represented by the string operator
\begin{equation}
  \mu_i \propto \prod_{m=-\infty}^{i-1} X_m \, .
  \label{eq:of-disorder-field}
\end{equation}

\subsection{SPT entangler and cluster OF model}

The cluster OF model is obtained by applying the SPT entangler, which can be implemented as the finite-depth unitary circuit
\begin{equation}
  \begin{aligned}
    U_{\rm SPT} & = \prod_i {\rm CZ}_{i,i+1} \, , \\
    {\rm CZ}_{i,i+1} & = \frac{1 + Z_i + Z_{i+1} - Z_iZ_{i+1}}{2} \, ,
  \end{aligned}
\end{equation}
where ${\rm CZ}$ is the controlled-$Z$ gate.
Its action on the Pauli operators is
\begin{equation}
  \begin{aligned}
    U_{\rm SPT}^{} X_i U_{\rm SPT}^{\dagger} & = Z_{i-1}X_iZ_{i+1} \, , \\
    U_{\rm SPT}^{} Z_i U_{\rm SPT}^{\dagger} & = Z_i \, .
  \end{aligned}
  \label{eq:cz-action}
\end{equation}
Under PBC, conjugation by the SPT entangler maps the original OF Hamiltonian to the cluster OF Hamiltonian
\begin{equation}
  \begin{aligned}
    H_{\rm cluster}^{} = {} & - \sum_i \left( Z_iZ_{i+1} + Z_iX_{i+1}Z_{i+2} \right) \\
                            & + g \sum_i \left( Z_iX_{i+1}Z_{i+3} + Z_iX_{i+2}Z_{i+3} \right) \, ,
  \end{aligned}
  \label{eq:model}
\end{equation}
which satisfies
\begin{equation}
  H_{\rm cluster}^{\rm PBC} = U_{\rm SPT}^{} H_{\rm OF}^{\rm PBC} U_{\rm SPT}^{\dagger} \, .
  \label{eq:pbc-fdu-equivalence}
\end{equation}
The two PBC Hamiltonians are therefore exactly isospectral and have identical local bulk CFT data.
The nearest-neighbor Ising interaction is unchanged by the SPT entangler, whereas the transverse-field term and the two $g$-dependent three-spin interaction terms are mapped to their cluster-decorated counterparts.
Both the original and cluster OF Hamiltonians are invariant under the global $\mathbb{Z}_2$ spin-flip symmetry generated by $\prod_i X_i$, as well as the antiunitary spinless time-reversal symmetry $\mathbb{Z}_2^T$ generated by complex conjugation $K$.

\begin{figure}[t]
  \centering
  \includegraphics[width=1.0\linewidth]{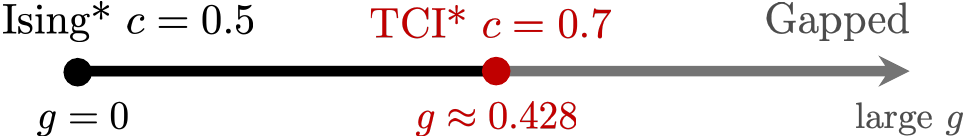}
  \caption{Ground-state phase diagram of the cluster OF model as a function of $g$. The symmetry-enriched Ising critical line with $c = 1/2$ terminates at the TCI* point near $g_c \approx 0.428$, beyond which the system enters a gapped regime.}
  \label{fig:cluster-phase-diagram}
\end{figure}

The same transformation gives the cluster-model representatives of the energy and fermionic fields,
\begin{equation}
  \begin{split}
    \epsilon_i & \propto \frac{1}{2} ( Z_{i-1}Z_i + Z_iZ_{i+1} ) - Z_{i-1}X_iZ_{i+1} \, , \\
    \psi_i & \propto \left[ \prod_{m=-\infty}^{i-2} X_m \right] ( X_{i-1}X_iZ_{i+1} + Y_{i-1} ) \, .
  \end{split}
  \label{eq:cluster-lattice-fields}
\end{equation}
Since the SPT entangler leaves $Z_i$ unchanged, the spin field is again represented by $Z_i$.
By contrast, it maps the disorder string to the SPT-dressed representative
\begin{equation}
  \tilde{\mu}_i \propto \Bigg[ \prod_{m=-\infty}^{i-2} X_m \Bigg] Y_{i-1}Z_i \, .
  \label{eq:cluster-disorder-field}
\end{equation}
The two disorder representatives have the same scaling dimension $\Delta_{\mu} = 3/40$ in their corresponding models but opposite charges under the spinless time-reversal symmetry: $K \mu K = \mu$ and $K \tilde{\mu} K = - \tilde{\mu}$.
At $g = 0$, the cluster OF model realizes the symmetry-enriched Ising, or Ising*, critical point with central charge $c = 1/2$~\cite{Verresen2021PRX,Yu2022PRL}.
The $g$-dependent interaction is irrelevant at this fixed point, and the Ising* critical line persists to finite $g$ and terminates at the symmetry-enriched TCI, which we call topological TCI or TCI*.
Again, the asterisk labels the symmetry enrichment rather than a different bulk TCI CFT.
This evolution is summarized in the ground-state phase diagram shown in Fig.~\ref{fig:cluster-phase-diagram}.
Our numerical verifications of the phase diagram and the emergent superconformal symmetry in the cluster OF model are presented in Appendix~\ref{app:bulk-susy}.

\subsection{Natural open-boundary terminations}

Under OBC, the natural termination of each Hamiltonian is obtained by retaining only terms whose support lies within the sites $1, \ldots, L$.
For the original OF model, this gives
\begin{equation}
  \begin{aligned}
    H_{\rm OF}^{\rm OBC} = {} & - \sum_{i=1}^{L} X_i - \sum_{i=1}^{L-1} Z_iZ_{i+1} \\
                              & + g \sum_{i=1}^{L-2} \left( X_iZ_{i+1}Z_{i+2} + Z_iZ_{i+1}X_{i+2} \right) \, ,
  \end{aligned}
  \label{eq:of-obc}
\end{equation}
whereas the natural cluster termination is
\begin{equation}
  \begin{aligned}
  H_{\rm cluster}^{\rm OBC} = {} & - \sum_{i=1}^{L-1} Z_iZ_{i+1} - \sum_{i=1}^{L-2} Z_iX_{i+1}Z_{i+2} \\
                                 & + g \sum_{i=1}^{L-3} \left( Z_iX_{i+1}Z_{i+3} + Z_iX_{i+2}Z_{i+3} \right) \, .
  \end{aligned}
  \label{eq:cluster-obc}
\end{equation}
For the open-chain circuit $U_{\rm SPT}^{\rm OBC} = \prod_{i=1}^{L-1} {\rm CZ}_{i,i+1}$, direct conjugation instead gives
\begin{equation}
  U_{\rm SPT}^{\rm OBC} H_{\rm OF}^{\rm OBC} (U_{\rm SPT}^{\rm OBC})^\dagger = H_{\rm cluster}^{\rm OBC} + H_{\rm end} \, ,
\end{equation}
with
\begin{equation}
  H_{\rm end} = - X_1Z_2 - Z_{L-1}X_L + g \left( X_1Z_3 + Z_{L-2}X_L \right) \, .
  \label{eq:obc-endpoint-terms}
\end{equation}
Thus, under OBC, the equivalence does not extend to the natural boundary terms: the natural terminations of the original and cluster OF chains are not mapped into one another.
This boundary mismatch helps explain, at the microscopic level, why the two models can exhibit distinct boundary physics under OBC.
The supersymmetry structure of the OF model under OBC is also discussed in Appendix~\ref{app:obc-susy}.

\subsection{Brief review of TCI BCFT}
\label{sec:tci-bcft-review}

To identify the conformal boundary conditions realized respectively by the two open chains, we first briefly review the basics of TCI BCFT needed here~\cite{cardy2008boundaryconformalfieldtheory,francesco2012conformal,zou2020prb}.

For TCI CFT, the six elementary Cardy boundary conditions can be labeled by the primary $\phi$ in Table~\ref{tab:tci-bulk-primaries}
\begin{equation}
  \begin{array}{@{}l@{\qquad}l@{}}
    |+\rangle  = |I\rangle \, , &
    |-\rangle  = |\epsilon''\rangle \, , \\
    |0+\rangle = |\epsilon\rangle \, , &
    |0-\rangle = |\epsilon'\rangle \, , \\
    |0\rangle  = |\sigma'\rangle \, , &
    |d\rangle  = |\sigma\rangle \, .
  \end{array}
  \label{eq:tci-cardy-boundaries}
\end{equation}
Here $|+\rangle$ and $|-\rangle$ are the two fixed boundaries, $|0+\rangle$ and $|0-\rangle$ are the two mixed boundaries, $|0\rangle$ is the free boundary, and $|d\rangle$ is the degenerate boundary.

At a conformal boundary fixed point, the boundary operator content organizes the universal low-energy spectrum by specifying which conformal towers appear and how many times each occurs.
For a Cardy boundary condition labeled by a primary field $\phi$, the boundary operator content is determined by the fusion rule $\phi \times \phi$.
Some useful TCI fusion rules are
\begin{equation}
  \begin{array}{@{}l@{\qquad}l@{}}
    I \times \phi = \phi \, , &
    \epsilon \times \epsilon = I + \epsilon' \, , \\
    \epsilon' \times \epsilon' = I + \epsilon' \, , &
    \epsilon'' \times \epsilon'' = I \, , \\
    \sigma \times \sigma = I + \epsilon + \epsilon' + \epsilon'' \, , &
    \sigma' \times \sigma' = I + \epsilon'' \, .
  \end{array}
\label{eq:tci-useful-fusion-rules}
\end{equation}

In a symmetry-preserving system, the two symmetry-related fixed sectors can also be combined into the spontaneously fixed boundary condition~\cite{Iino2020PRB}
\begin{equation}
  |\!+\!\&-\rangle = |+\rangle + |-\rangle \, .
  \label{eq:sf-boundary}
\end{equation}
The microscopic Hamiltonian preserves the global $\mathbb{Z}_2$ symmetry, which exchanges the two fixed sectors, while in the thermodynamic limit, an infinitesimal boundary field can select either fixed polarization.
Hereafter, we also use $|\mathrm{f}\rangle$ and $|\mathrm{s}\rangle$ to denote the free and spontaneously fixed boundary conditions, respectively.

When both ends have free boundary conditions, the boundary operator content follows from the fusion rule
\begin{equation}
  [\sigma'] \times [\sigma'] = [I] + [\epsilon''] \, .
  \label{eq:free-open-spectrum}
\end{equation}
For degenerate boundary conditions at both ends, one similarly obtains
\begin{equation}
  [\sigma] \times [\sigma] = [I] + [\epsilon] + [\epsilon'] + [\epsilon''] \, .
  \label{eq:degenerate-open-spectrum}
\end{equation}
For spontaneously fixed boundary conditions at both ends,
\begin{equation}
  ( [I] + [\epsilon''] ) \times ( [I] + [\epsilon''] ) = 2[I] + 2[\epsilon''] \, .
  \label{eq:sf-open-spectrum}
\end{equation}
In our work, the two identity towers (i.e., $2[I]$) correspond to the two symmetry-related ground-state sectors with parallel endpoint polarizations, whereas the two $[\epsilon'']$ towers correspond to opposite endpoint polarizations, which is similar to the ferromagnetic Ising case~\cite{Yu2022PRL}.

\section{Conformal boundary condition of the cluster OF model under OBC}
\label{sec:natural-boundary}

Having established that the original and cluster OF models share the same bulk TCI CFT but have inequivalent natural boundary terms under OBC, we now determine the conformal boundary conditions realized by their natural open-chain terminations.
In particular, we examine whether the natural termination of the cluster OF chain realizes the spontaneously fixed boundary condition, in contrast to the free boundary condition of the original OF model.
We address this question using three complementary diagnostics: spontaneous boundary magnetization, boundary-bulk correlations, and the low-energy open-chain spectrum.
Details of the DMRG calculations are provided in Appendix~\ref{app:numerical-methodology}.

\subsection{Spontaneous boundary magnetization}

\begin{figure}[t]
  \centering
  \includegraphics[width=0.9\linewidth]{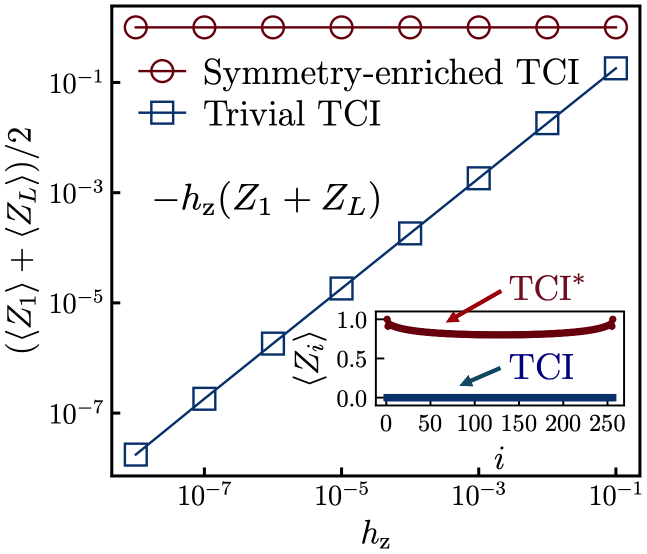}
  \caption{Boundary magnetization $m_{\rm bdy} = [ \langle Z_1 \rangle + \langle Z_L \rangle ] / 2$ versus $h_\text{z}$ with the boundary-field term $H_\text{bdy} = - h_\text{z} ( Z_{1} + Z_{L} )$ for the original (blue square) and cluster (red circle) OF models at the tricritical point. Inset: the spatial profile of the local magnetization $\langle Z_{i} \rangle$ across the chain under a vanishingly small $h_\text{z} = 10^{-8}$ for $L = 256$. Simulations are performed under OBC.}
  \label{fig:boundary-magnetization}
\end{figure}

We begin by examining the spontaneous boundary magnetization.
To this end, we apply a weak symmetry-breaking boundary field $H_{\rm bdy} = - h_\text{z} ( Z_1 + Z_L )$ and measure the boundary magnetization $m_{\rm bdy} = [ \langle Z_1 \rangle + \langle Z_L \rangle ] / 2$.
As shown in Fig.~\ref{fig:boundary-magnetization}, the boundary magnetization of the cluster OF model remains finite as $h_\text{z}$ is reduced.
By contrast, the boundary magnetization of the original OF model vanishes as $h_\text{z} \to 0$.
These contrasting behaviors provide initial evidence that the natural terminations of the cluster and original OF chains realize the spontaneously fixed and free boundary conditions, respectively.

The inset further compares the magnetization profiles of the cluster and original OF models under a small boundary field $h_\text{z} = 10^{-8}$ for a finite chain of size $L = 256$.
For the cluster OF model, the boundary field selects a polarized boundary sector and induces a nonzero magnetization throughout the finite critical chain.
This bulk magnetization is a finite-size response and vanishes in the thermodynamic limit;
in particular, the magnetization at the center of the chain should scale as $\langle Z_{L/2} \rangle \sim L^{-\Delta_\sigma}$, where $\Delta_\sigma = 3/40$ is the bulk spin scaling dimension.
By contrast, the magnetization of the original OF model remains negligible throughout the chain, consistent with the absence of spontaneous boundary magnetization.

\subsection{Boundary-bulk correlations}
\label{sec:boundary-correlations-boe}

\begin{figure*}[t]
  \centering
  \includegraphics[width=0.8\linewidth]{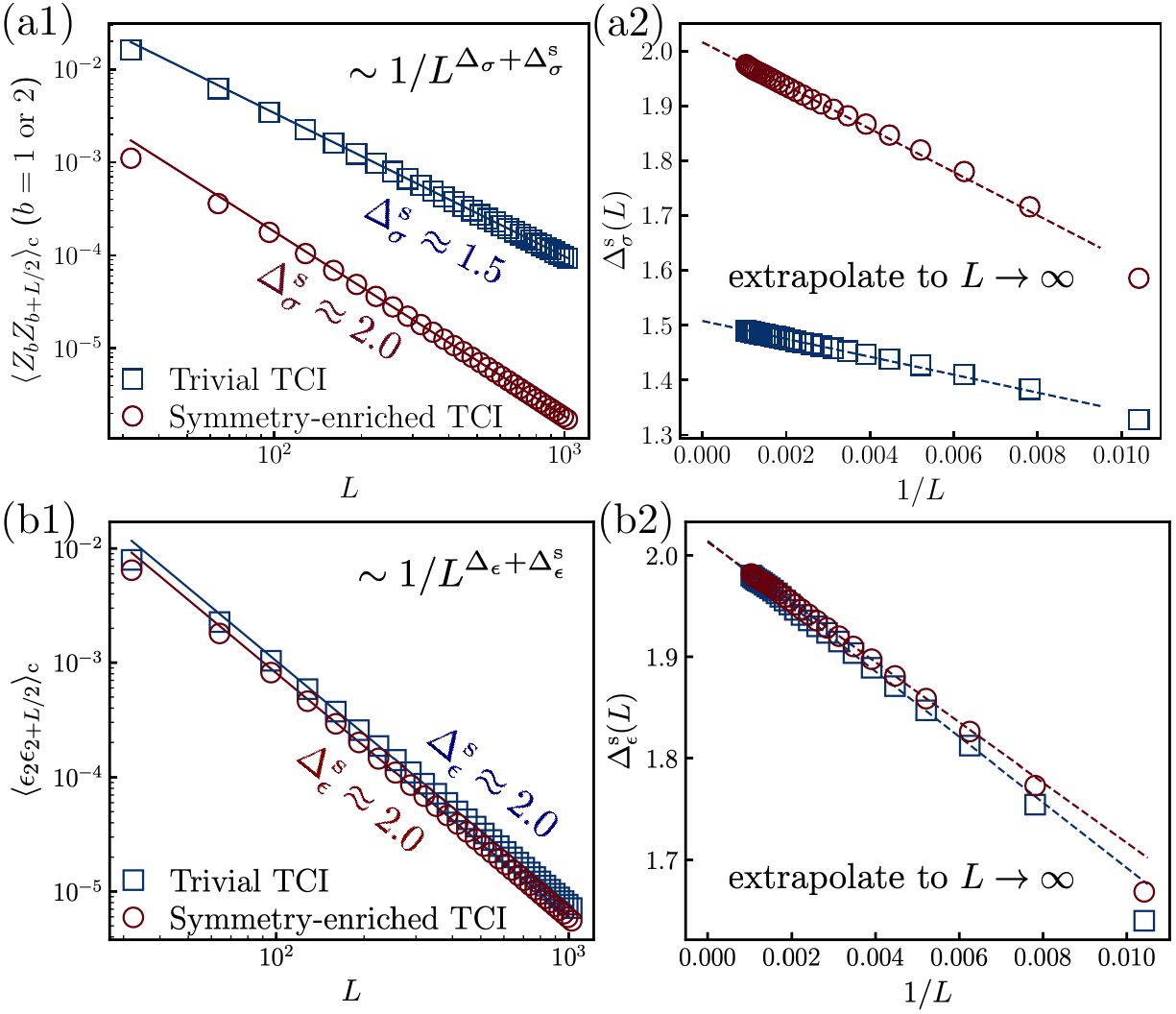}
  \caption{(a1) Finite-size scaling of the boundary-bulk connected spin-spin correlation function, which scales as $\sim L^{-(\Delta_{\sigma} + \Delta_{\sigma}^\text{s})}$. Here we have chosen $r = L/2$ for finite-size systems of size $L$. Taking the known bulk scaling dimension $\Delta_{\sigma} = 3/40$, we extract $\Delta_{\sigma}^\text{s} \approx 1.5$ ($\Delta_{\sigma}^\text{s} \approx 2.0$) for the original (cluster) OF model. (a2) shows the finite-size exponent $\Delta_{\sigma}^\text{s}(L)$ obtained via power-law fits using data from system sizes $L$, $L \pm 32$, and $L \pm 64$. (b1) Finite-size scaling of the boundary-bulk connected energy-energy correlation function, which scales as $\sim L^{-(\Delta_{\epsilon} + \Delta_{\epsilon}^\text{s})}$. Fixing the bulk scaling dimension to the analytical value $\Delta_{\epsilon} = 1/5$, we extract the boundary scaling dimensions $\Delta_{\epsilon}^\text{s} \approx 2.0$ for both the original and cluster OF models. (b2) The finite-size exponent $\Delta_{\epsilon}^\text{s}(L)$ obtained via power-law fits using data from system sizes $L$, $L\pm32$, and $L\pm64$, which extrapolates to $\Delta_{\epsilon}^\text{s} \approx 2.01$ in the thermodynamic limit. Simulations are performed under OBC.}
  \label{fig:boundary-correlations}
\end{figure*}

The spontaneous boundary magnetization provides a direct distinction between the boundary responses of the original and cluster OF models.
To characterize the corresponding boundary fixed points more quantitatively, we study connected boundary-bulk correlations of the spin and energy fields.
For a bulk primary field $\phi$ with bulk scaling dimension $\Delta_\phi$, its connected boundary-bulk correlation considered below is governed by $\Delta_\phi + \Delta_\phi^{\rm s}$, where $\Delta_\phi^{\rm s}$ is the scaling dimension of the leading boundary operator with a nontrivial contribution to the correlation function.
Through the boundary operator expansion (BOE), this boundary exponent $\Delta_\phi^{\rm s}$ depends on both the bulk field and the conformal boundary condition~\cite{cardy2008boundaryconformalfieldtheory}.
For the cluster OF model, the twofold open-chain ground-state manifold contains two symmetry-related boundary-polarized sectors.
In the correlation-function calculations below, we choose one of these two ground states.
Equivalently, an infinitesimal boundary field can be used to lift this twofold degeneracy and select a definite boundary-polarized sector.

We first consider the spin field $\sigma_i \propto Z_i$.
For a boundary spin at site $b$ and a bulk spin at a distance $r$ from it, the connected correlation scales as
\begin{equation}
  \langle Z_b Z_{b+r} \rangle_\text{c} \propto r^{-(\Delta_\sigma+\Delta_\sigma^\text{s})} \, .
\end{equation}
For a selected polarized ground state of the cluster OF chain, the outermost spin $Z_1$ is fixed to $+1$ or $-1$, so its connected correlation function with any other spin vanishes.
We therefore take $b = 2$ for the cluster OF model and $b = 1$ for the original OF model~\cite{Yu2022PRL}.
For the finite-size scaling, we set $r = L/2$.
As shown in Fig.~\ref{fig:boundary-correlations}(a1), we obtain
\begin{equation}
  \begin{split}
    \Delta_\sigma^{\rm s} = 1.5 \, , \ \text{TCI QCP} \, , \qquad
    \Delta_\sigma^{\rm s} = 2.0 \, , \ \text{TCI* QCP} \, .
  \end{split}
\end{equation}
The finite-size extrapolations of the boundary exponents are shown in Fig.~\ref{fig:boundary-correlations}(a2).
The different values of $\Delta_\sigma^{\rm s}$ provide a second indication that the two natural terminations realize distinct conformal boundary conditions.

We next examine the boundary-bulk correlation of the energy field $\epsilon$ using the corresponding lattice representatives given in Eqs.~\eqref{eq:of-lattice-fields} and~\eqref{eq:cluster-lattice-fields}.
Under OBC, $\epsilon_1$ is not supported by either lattice representative, so we use $\epsilon_2$ as the boundary energy-field insertion.
For an energy operator near the boundary and another at a distance $r$ from it, the connected correlation scales as
\begin{equation}
  \langle \epsilon_{2} \epsilon_{2+r} \rangle_\text{c} \propto r^{-(\Delta_{\epsilon} + \Delta_{\epsilon}^\text{s})} \, ,
\end{equation}
where $\Delta_{\epsilon} = 1/5$ is the bulk scaling dimension and $\Delta_{\epsilon}^{\rm s}$ is the boundary exponent to be extracted.
Figure~\ref{fig:boundary-correlations}(b1) shows the finite-size scaling of $\langle \epsilon_{2} \epsilon_{2+L/2}\rangle_\text{c}$ for the original and cluster OF models, while Fig.~\ref{fig:boundary-correlations}(b2) shows the corresponding finite-size exponents.
In contrast to the spin field, the energy field has the same boundary scaling dimension in the two models, i.e., $\Delta_{\epsilon}^{\rm s} \approx 2.0$\,.

The numerical exponents extracted above can be understood using the BOE analyses.
For a Cardy boundary state $|a\rangle$, the candidate boundary primary fields are constrained by the boundary operator spectrum in the $a$-$a$ sector, obtained from the fusion rule $a \times a$ reviewed in Sec.~\ref{sec:tci-bcft-review}.
For the diagonal bulk primaries considered here, a boundary field appearing in the BOE must also be compatible with the bulk fusion rule $\phi \times \phi$.

For the free boundary condition realized by the original OF model, the boundary spectrum is $[\sigma'] \times [\sigma'] = [I] + [\epsilon'']$.
Since the bulk spin field obeys $[\sigma] \times [\sigma] = [I] + [\epsilon] + [\epsilon'] + [\epsilon'']$, the boundary primary $[\epsilon'']$ can appear in its BOE.
The leading nontrivial boundary primary therefore has scaling dimension $\Delta_{\epsilon''}^{\rm bdy} = 3/2$.
This gives $\Delta_\sigma^{\rm s} = 3/2$ for the original OF model.

For the spontaneously fixed boundary condition realized by the cluster OF model, the correlation calculation probes a single polarized fixed sector.
The corresponding same-sector boundary spectrum contains only the identity family $[I]$, since $[I] \times [I] = [\epsilon''] \times [\epsilon''] = [I]$.
After the identity contribution is removed from the connected correlation, the leading nontrivial contribution is the boundary stress tensor with scaling dimension $\Delta_T=2$.
This explains the observed boundary exponent $\Delta_\sigma^{\rm s} = 2$ for the cluster OF model.

For the energy field, we have $[\epsilon] \times [\epsilon] = [I] + [\epsilon']$.
However, the boundary primary $[\epsilon']$ is absent from both the free boundary spectrum $([I] + [\epsilon''])$ and the same-sector fixed boundary spectrum $([I])$.
Consequently, after subtracting the identity contribution, the leading nontrivial term is again the boundary stress tensor in both cases.
This accounts for the same value $\Delta_\epsilon^{\rm s} = 2$ extracted from the two models.
For comparison, the corresponding BOE analysis for the critical transverse-field Ising and cluster Ising chains is given in Appendix~\ref{app:ising-boe}.

The associated bulk and boundary scaling dimensions are summarized in Table~\ref{tab:scaling_comparison}.

\begin{table}[b]
  \caption{Bulk and boundary scaling dimensions of the spin $\sigma$ and energy $\epsilon$ operators for the trivial TCI (obtained from the original OF model with free boundary condition) and the topological TCI (obtained from the cluster OF model with spontaneously fixed boundary condition).}
  \label{tab:scaling_comparison}
  \begin{ruledtabular}
    \begin{tabular}{lcccc}
    Class \textbackslash\, Scaling Dimensions & $\Delta_\sigma$ & $\Delta_\sigma^\text{s}$ & $\Delta_\epsilon$ & $\Delta_\epsilon^\text{s}$ \\
    \hline
    Trivial TCI (free b.c.) & $3/40$ & $1.5$ & $1/5$ & $2.0$ \\
    Topological TCI (spont. fixed b.c.) & $3/40$ & $2.0$ & $1/5$ & $2.0$ \\
    \end{tabular}
  \end{ruledtabular}
\end{table}

\subsection{Low-energy open-chain spectra}

\begin{figure}[t]
  \centering
  \includegraphics[width=0.9\linewidth]{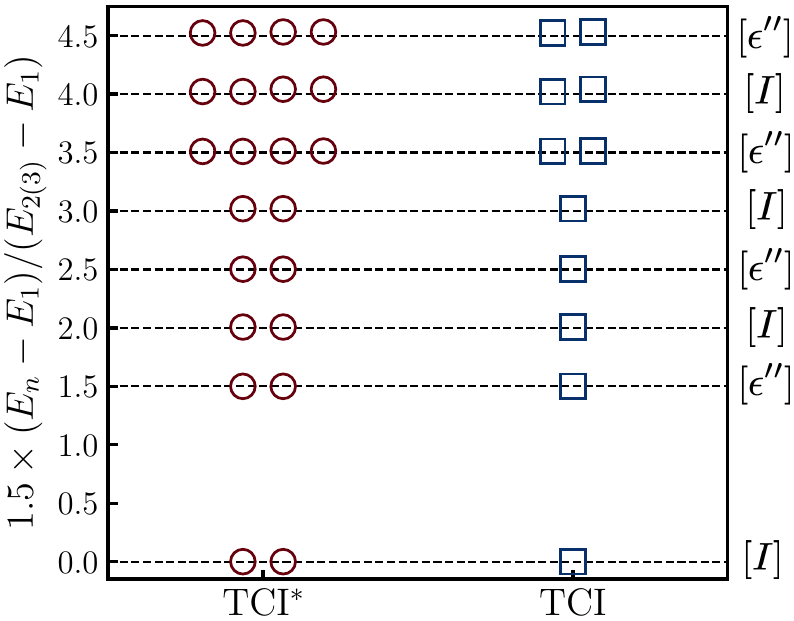}
  \caption{Rescaled low-energy open-chain spectra of the cluster (left) and original (right) OF models at the tricritical point for $L = 128$. The spectra are consistent with the BCFT operator content of the spontaneously fixed and free boundary conditions, respectively. Simulations are performed under OBC.}
  \label{fig:boundary-spectrum}
\end{figure}

To further identify the realized conformal boundary conditions, we finally examine the low-energy spectra under OBC.
Figure~\ref{fig:boundary-spectrum} compares the low-energy spectra of the cluster and original OF models at their natural open-chain terminations.
The rescaled spectrum of the original OF model matches the boundary operator content $[I] + [\epsilon'']$, corresponding to the free boundary condition $|\mathrm{f}\rangle = |\sigma'\rangle$.
In contrast, the cluster OF model exhibits two identical copies, $2[I] + 2[\epsilon'']$, matching the spontaneously fixed boundary condition $|\mathrm{s}\rangle = |\!+\!\&-\rangle$, whose boundary operator content is obtained from $([I] + [\epsilon'']) \times ([I] + [\epsilon'']) = 2[I] + 2[\epsilon'']$;
see Sec.~\ref{sec:tci-bcft-review} for the TCI Cardy states and fusion rules~\cite{francesco2012conformal,Iino2020PRB,Ian_Affleck_2000,CHIM1996IJMPA}.
For the cluster OF model, the factor of two reflects the edge-mode degeneracy, giving the topological twofold degeneracy at the $\mathrm{TCI}^*$ QCP, protected by $\mathbb{Z}_2\times\mathbb{Z}_2^T$ symmetry~\cite{Verresen2021PRX}.

\section{Finite-Size Splitting of the Boundary Degeneracy in the Cluster OF Model}
\label{sec:finite-size-splitting}

\begin{figure}[t]
  \centering
  \includegraphics[width=0.9\linewidth]{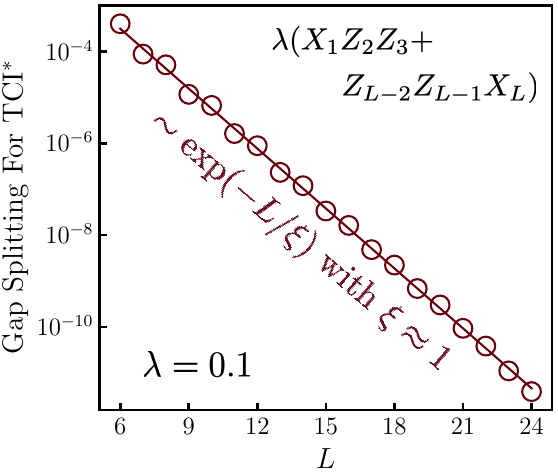}
  \caption{Finite-size energy splitting of the twofold boundary degeneracy at TCI* QCP of the cluster OF model under the representative $\mathbb{Z}_2 \times \mathbb{Z}_2^T$-preserving boundary perturbation $\lambda( X_1Z_2Z_3 + Z_{L-2}Z_{L-1}X_L )$ with $\lambda = 0.1$\,. The energy splitting is numerically consistent with exponential decay over the displayed system sizes. Simulations are performed under OBC using exact diagonalization.}
  \label{fig:representative-exp-splitting}
\end{figure}

At the natural open-chain termination of the cluster OF model, we have shown that the spontaneously fixed boundary condition supports a topological twofold degeneracy.
The two boundary-polarized ground states, $|\!\!\uparrow_L\uparrow_R\rangle$ and $|\!\!\downarrow_L\downarrow_R\rangle$, where the arrows denote the magnetization directions at the two ends, are related by the global spin-flip symmetry.
A boundary term that preserves the protecting $\mathbb{Z}_2 \times \mathbb{Z}_2^T$ symmetry can couple these edge modes through the critical bulk and lift their exact microscopic degeneracy at finite size.
We first consider the representative boundary perturbation $\lambda ( X_1 Z_2 Z_3 + Z_{L-2} Z_{L-1} X_L )$ with $\lambda = 0.1$\,.
This is exactly the same symmetry-preserving boundary perturbation considered in the Ising* case~\cite{Verresen2021PRX}, here applied to the cluster OF model at the TCI* QCP.
As shown in Fig.~\ref{fig:representative-exp-splitting}, the resulting finite-size energy splitting is consistent with
\begin{equation}
  \delta E(L) \sim \text{e}^{-L/\xi} \, .
  \label{eq:exponential-splitting}
\end{equation}

\begin{figure*}[t]
  \centering
  \includegraphics[width=0.8\textwidth]{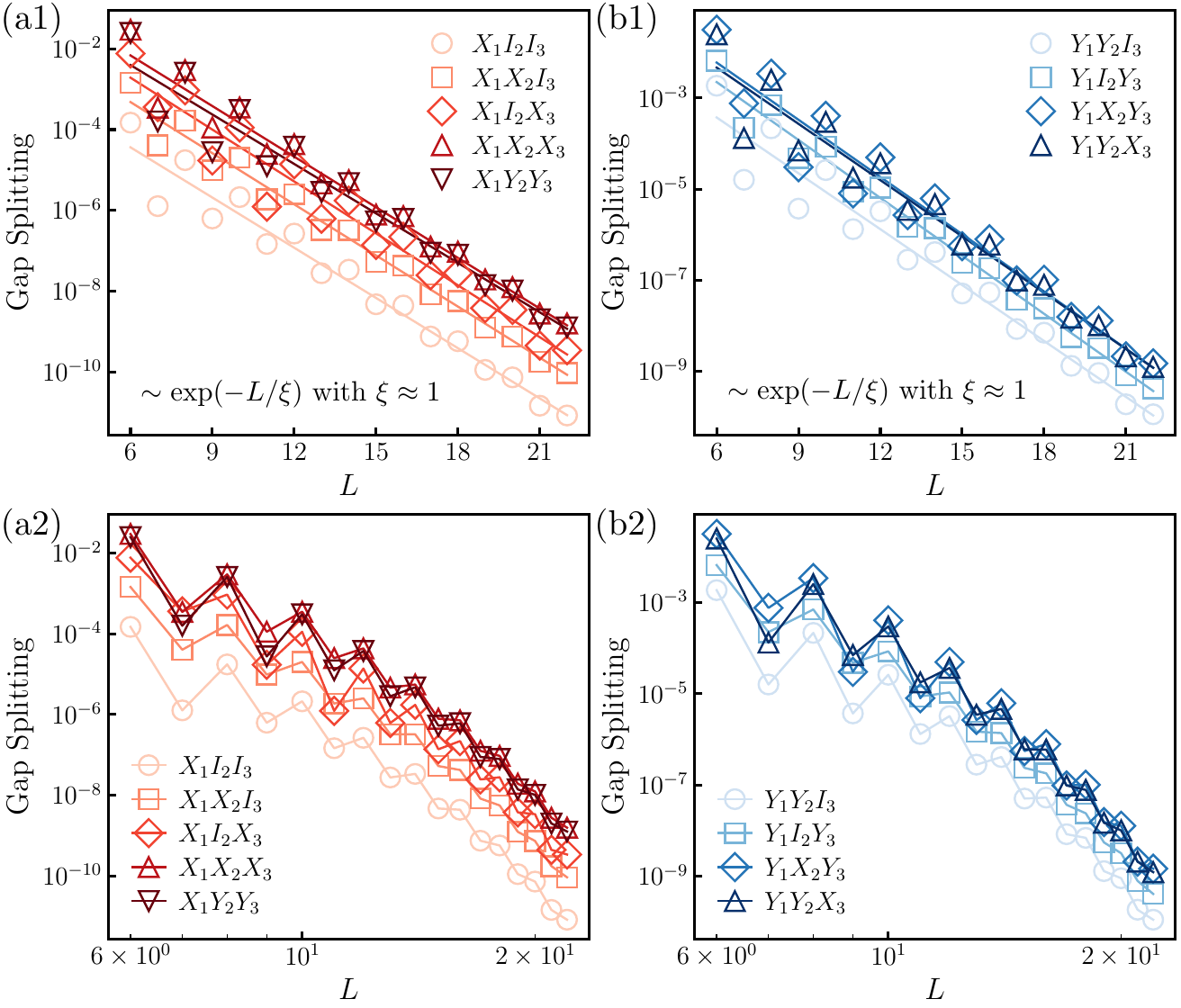}
  \caption{Robustness of the twofold degeneracy at TCI* against symmetry-preserving boundary perturbations. We compute the gap splitting of the ground-state manifold induced by all possible three-site Pauli-string-type perturbations at the boundaries, $\lambda (S_{1}S_{2}S_{3} + S_{L-2}S_{L-1}S_{L})$, where $S_{i} \in \{ I_{i}, X_{i}, Y_{i}, Z_{i} \}$ (with $S_{L+1-i} = S_{i}$) and $\lambda = 0.1$, that strictly respect both the global $\mathbb{Z}_{2}$ spin-flip and antiunitary time-reversal symmetries. The left and right panels display the results for perturbations where the outermost operator is $X_{1}$ and $Y_{1}$, respectively. The comparison between log-linear and log-log plots consistently reveals a rapid exponential decay $\sim \exp(-L/\xi)$ with $\xi \approx 1$, ruling out power-law decaying behavior. Perturbations initiating with $I_{1}$ and $Z_{1}$ are omitted as they maintain the exact twofold degeneracy. The specific $X_{1}Z_{2}Z_{3}$ case has been analyzed in Fig.~\ref{fig:representative-exp-splitting}. Simulations are performed under OBC using exact diagonalization.}
  \label{fig:systematic-exp-splitting}
\end{figure*}

The TCI* and Ising* QCPs are both protected by the $\mathbb{Z}_2 \times \mathbb{Z}_2^T$ symmetry~\cite{Verresen2021PRX}.
In the Ising* case, the corresponding symmetry-preserving finite-size splitting is algebraic, $\delta E(L) \sim L^{-14}$.
Following the same character-based argument of Ref.~\cite{Verresen2021PRX}, one would similarly anticipate a power-law splitting at the TCI* QCP, leading to the naive estimate $\delta E(L) \sim L^{-12}$ (see Appendix~\ref{app:splitting-analysis} for details).
The exponential behavior in Fig.~\ref{fig:representative-exp-splitting} is therefore qualitatively different from both the Ising* result and the naive power-law expectation.

One possible explanation is that the ordinary Virasoro characters specify the boundary operator content but do not resolve the $\mathbb{Z}_2^T$ charge of individual operators.
An operator present in an ordinary character may therefore still be forbidden from coupling the two edges by the microscopic antiunitary symmetry.
This is not a derivation of Eq.~\eqref{eq:exponential-splitting};
the symmetry-resolved character issue and the resulting open problem are discussed in Appendix~\ref{app:splitting-analysis}.

To exclude the possibility that the representative $XZZ$ perturbation is exceptional, we scan $\mathbb{Z}_2 \times \mathbb{Z}_2^T$-preserving Pauli-string boundary terms.
Figure~\ref{fig:systematic-exp-splitting} displays the complete enumerated set with support fewer than $4$ that can lift the exact degeneracy. 
In practice, our tests have even extended to Pauli strings with support fewer than $7$.
For all tested symmetry-preserving boundary terms that lift the exact degeneracy, the finite-size splitting is numerically consistent with exponential decay over the accessible system sizes and numerical resolution.
Thus, within the perturbation families and system sizes examined here, the topological twofold edge modes found at TCI* show stronger finite-size robustness than the power-law-split edge modes of Ising*.
This qualitative difference makes the cluster-OF realization of topological TCI a valuable example for the study of gapless SPT phases~\cite{Li2025SciPost}.

\section{Symmetry-Enriched TCI Spontaneously Fixed Boundary Condition}
\label{sec:symmetry-enriched-boundaries}

So far, the distinct boundary behavior of the two OF models appears to originate from the different conformal boundary conditions they realize.
However, in the original OF model, the spontaneously fixed boundary condition can also be reached by tuning an appropriate boundary term from its natural free boundary condition~\cite{Iino2020PRB}.
This raises a natural question: are the two realizations of the spontaneously fixed boundary condition, one arising from symmetry enrichment and the other from boundary RG flow, physically equivalent or fundamentally distinct?
To address this question, we consider boundary deformations that preserve the global $\mathbb{Z}_2 \times \mathbb{Z}_2^T$ symmetry.

\subsection{Boundary RG flows and stability under symmetry-preserving boundary deformations}
\label{sec:boundary-deformations}

\begin{figure*}[t]
  \centering
  \includegraphics[width=1.0\textwidth]{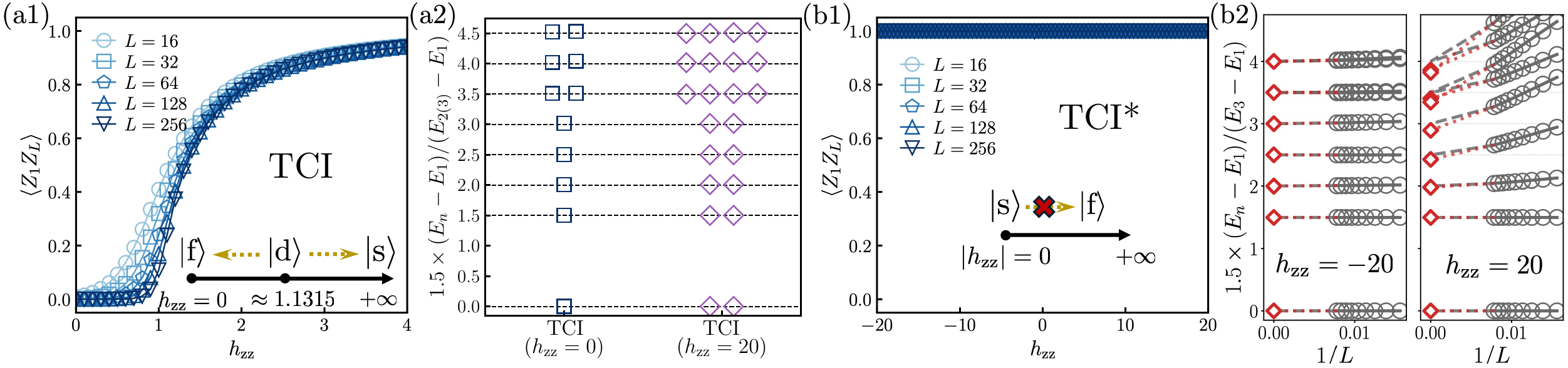}
  \caption{Boundary spin-spin correlation $\langle Z_1 Z_L \rangle$ as a function of $h_{\rm zz}$ under $H_{\rm bdy} = - h_{\rm zz} ( Z_1 Z_2 + Z_{L-1} Z_L )$ for (a1) the original and (b1) cluster OF models at the tricritical point. The inset in (a1) shows the corresponding boundary RG flows among the free $|\mathrm{f}\rangle$, degenerate $|\mathrm{d}\rangle$, and spontaneously fixed $|\mathrm{s}\rangle$ boundary conditions. (a2) displays the rescaled low-energy spectrum $1.5(E_n-E_1)/(E_{2(3)}-E_1)$ of the original OF model at $h_{\rm zz} = 0$ ($h_{\rm zz} = 20$), which is consistent with the BCFT operator content of the free (spontaneously fixed) boundary condition. The inset of (b1) indicates that the spontaneously fixed boundary condition of the cluster OF chain can not be driven to the free one via the same $ZZ$-type boundary deformation. (b2) shows the rescaled spectrum $1.5(E_n-E_1)/(E_{3}-E_1)$ at $h_{\rm zz} = \pm 20$, which is consistent within numerical resolution with the BCFT operator content of $|\mathrm{s}\rangle$. Red dashed lines show finite-size extrapolations, while red diamonds indicate extrapolated thermodynamic-limit values (see Appendix~\ref{app:boundary-perturbations} for details); gray dashed lines mark the corresponding scaling dimensions. The red-diamond levels have degeneracies $2,2,2,2,2,4,4$ from bottom to top. Simulations are performed under OBC.}
  \label{fig:boundary-rg-comparison}
\end{figure*}

We first show that the spontaneously fixed boundary condition can be reached in the original OF model by adding a $ZZ$-type boundary term $H_{\rm bdy} = - h_{\rm zz} ( Z_1 Z_2 + Z_{L-1} Z_L )$.
As shown in Fig.~\ref{fig:boundary-rg-comparison}(a1), the boundary correlation $\langle Z_1 Z_L \rangle$ grows continuously from zero as $h_{\rm zz}$ increases, indicating a boundary phase transition (see Appendix~\ref{app:boundary-rg} for the precise determination of $h_{\rm zz}^{c} \simeq 1.1315$).
Across this transition, the free boundary condition flows through an unstable degenerate boundary condition to the spontaneously fixed one~\cite{Iino2020PRB}.
Thus, in the original OF model, the spontaneously fixed boundary condition is accessible from the free boundary condition via boundary RG flow, which is also evidenced by the low-energy spectra shown in Fig.~\ref{fig:boundary-rg-comparison}(a2) for $h_{\rm zz} = 0$ and $h_{\rm zz} = 20$.
The transition is \textit{reversible}, with the two boundary conditions interconverted by tuning $h_{\rm zz}$ across $h_{\rm zz}^{c}$; see the inset of Fig.~\ref{fig:boundary-rg-comparison}(a1).
The gap-ratio calculation and Affleck--Ludwig boundary $g$-function analyses in Appendix~\ref{app:boundary-rg} provide further support for this boundary RG-flow structure.

In the TCI boundary RG diagram, the spontaneously fixed boundary condition is actually a \textit{stable} RG fixed point along the $\mathbb{Z}_2$-preserving path with no relevant symmetric boundary perturbation.
This implies its stability against weak $\mathbb{Z}_2$-symmetry-preserving perturbations, but does not by itself determine the effect of a finite or strong microscopic boundary deformation.
The RG trajectory of the original OF model shown above provides an example: a finite $ZZ$-type boundary term tunes the free boundary condition through the unstable degenerate fixed point to the spontaneously fixed one and vice versa.
Here, it is instructive to compare the situation with the Ising case.
In the transverse-field Ising chain, realizing the spontaneously fixed boundary condition may require taking the strength of the $ZZ$-type boundary deformation to infinity, since the spontaneously fixed boundary condition corresponds to an unstable boundary RG fixed point there.

We can therefore ask whether the natural cluster realization of the spontaneously fixed boundary condition can likewise be tuned to the free one by a symmetry-preserving boundary deformation.
For \textit{a direct comparison}, we first apply the same $ZZ$-type boundary term.
In stark contrast, the boundary correlation $\langle Z_1 Z_L \rangle$ in the cluster OF model remains pinned at unity for all $h_{\rm zz}$; see Fig.~\ref{fig:boundary-rg-comparison}(b1).
This implies that the topological degeneracy remains intact throughout.
The low-energy spectrum at $h_{\rm zz} = \pm 20$, shown in Fig.~\ref{fig:boundary-rg-comparison}(b2), is also consistent within numerical resolution with the BCFT operator content of the spontaneously fixed boundary condition.

Interestingly, we further find an unexpected fourfold ground-state degeneracy window near $h_{\rm zz} \approx -1$.
The robustness under the $ZZ$-type boundary term is expected, since $Z_1$ and $Z_L$ commute with the deformed Hamiltonian for every $h_{\rm zz}$, protecting the twofold degeneracy by exact boundary conservation.
This conservation alone, however, does not account for the enhanced fourfold degeneracy.
Since a full characterization of this fourfold degeneracy requires extended numerical scans and a detailed analysis of the boundary RG structure, we present the corresponding numerical data and discussion in Appendix~\ref{app:boundary-perturbations}.
Thus, unlike in the original OF model, the $ZZ$-type boundary term does not convert the cluster OF model from the spontaneously fixed to the free boundary condition.
This already suggests that the two spontaneously fixed boundary conditions, realized through boundary RG flow and symmetry enrichment, are physically distinct.

\begin{figure*}[t]
  \centering
  \includegraphics[width=1.0\textwidth]{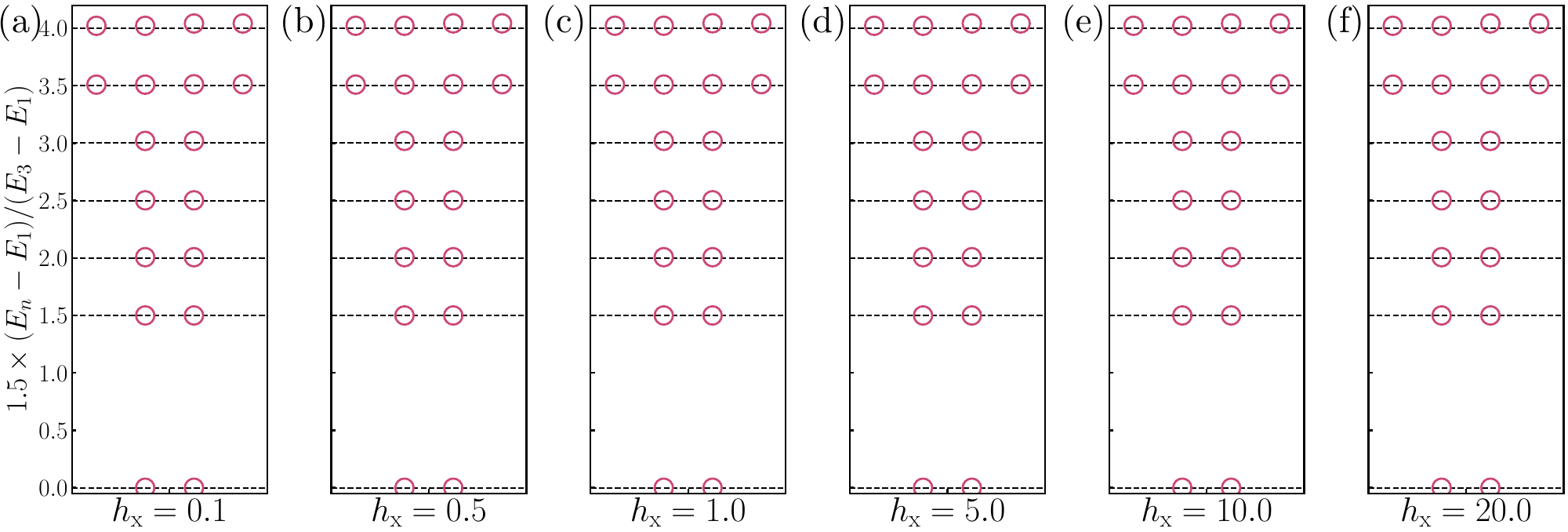}
  \caption{(a)-(f) Rescaled open-boundary energy spectra, $1.5 (E_{n} - E_{1}) / (E_{3} - E_{1})$, of the cluster OF model at the TCI* QCP in the presence of a boundary term $H_\text{bdy} = - h_\text{x} \left( X_{1} + X_{L} \right)$. The panels illustrate the spectral evolution across a wide range of finite perturbation strengths, specifically for $h_\text{x} = 0.1$, $0.5$, $1.0$, $5.0$, $10.0$, and $20.0$, respectively, from left to right. All the numerically extracted low-lying energy spectra (open circles) exhibit perfect agreement with the BCFT operator content (dashed lines) of the spontaneously fixed boundary condition. Simulated system size is $L = 128$ under OBC.}
  \label{fig:xstable}
\end{figure*}

We next perform a more stringent test by applying the symmetry-preserving boundary field $H_{\rm bdy} = - h_{\rm x} ( X_1 + X_L )$, which removes the exact conservation of the outer-boundary $Z$ spins.
One might expect a sufficiently large $h_{\rm x}$ to destroy the spontaneous boundary magnetization at the TCI* QCP.
Instead, for $h_{\rm x} = 0.1, 0.5, 1, 5, 10,$ and $20$, the twofold low-energy manifold remains intact and the rescaled spectrum remains consistent within numerical resolution with the spontaneously fixed BCFT operator content; see Fig.~\ref{fig:xstable}.

\begin{figure}[t]
  \centering
  \includegraphics[width=0.9\linewidth]{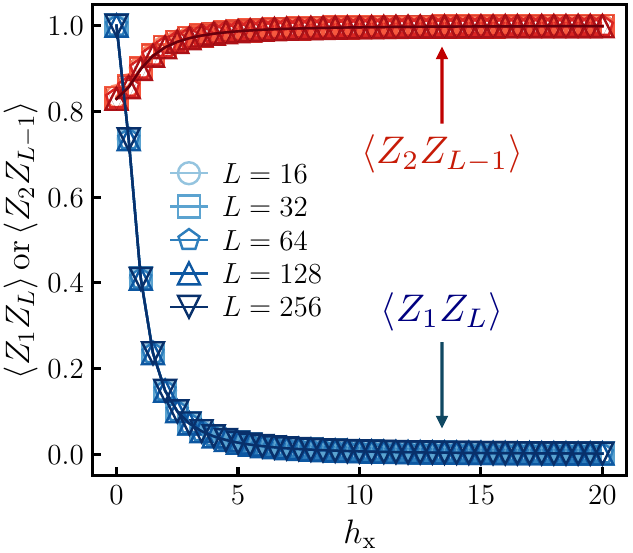}
  \caption{Boundary-boundary spin correlations $\langle{Z_{1}Z_{L}}\rangle$ and $\langle{Z_{2}Z_{L-1}}\rangle$ as a function of the boundary-field strength $h_\text{x}$ with $H_\text{bdy} = - h_\text{x} (X_{1} + X_{L})$ for the cluster OF model at TCI* QCP. Simulations are performed under OBC.}
  \label{fig:xstable_correlation}
\end{figure}

The persistence of the twofold degeneracy can be understood in the infinite-$h_{\rm x}$ limit.
In this limit, the boundary field locks the endpoint spins in their $X$-polarized states.
Projecting to this low-energy subspace suppresses boundary terms that do not commute with $X_1$ or $X_L$, and the remaining effective Hamiltonian describes a clean cluster OF chain on sites $2$ through $L-1$, together with two isolated endpoint spins polarized in the $X$ direction.
The spontaneous boundary magnetization is therefore not destroyed, but instead shifts inward to sites $2$ and $L-1$.
This mechanism is directly reflected in Fig.~\ref{fig:xstable_correlation}: the outer correlation $\langle Z_1 Z_L \rangle$ decreases with $h_{\rm x}$, whereas the inner correlation $\langle Z_2 Z_{L-1} \rangle$ increases toward unity.
Thus, the twofold boundary degeneracy remains stable under the $X$ boundary deformation from finite $h_{\rm x}$ to the strong-field limit, a result that cannot be inferred from the perturbative RG stability of the spontaneously fixed boundary condition alone.

Taken together, the $ZZ$ and $X$ results contrast with the boundary RG flow of the original OF model: within the symmetry-preserving boundary deformations studied here, we do not find a path that converts the natural cluster realization from the spontaneously fixed to the free boundary condition.
This difference supports the distinction between the symmetry-enriched realization of the spontaneously fixed boundary condition and the realization reached through boundary RG flow.
These observations do not establish stability under arbitrary symmetry-preserving boundary deformations;
a general analytic argument, or a numerical analysis covering all symmetry-preserving boundary deformations, remains an open problem.

\subsection{$\mathbb{Z}_2^T$ charges of disorder fields and nonlocal string operators}

\begin{figure*}[t]
  \centering
  \includegraphics[width=0.8\textwidth]{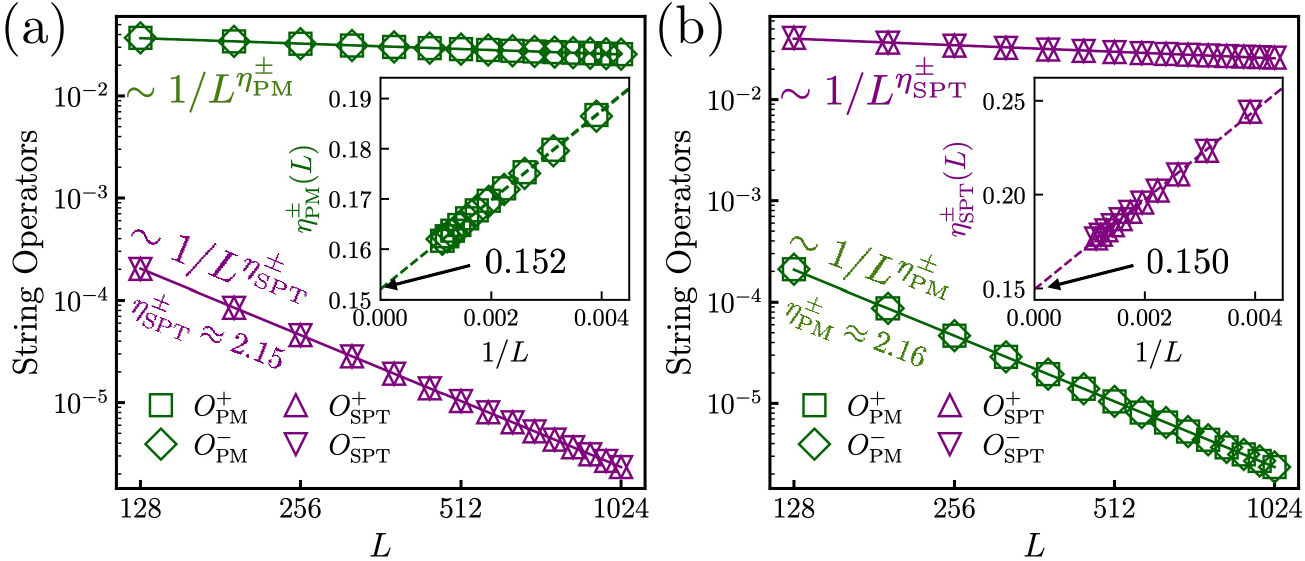}
  \caption{Log-log plots of the nonlocal string operators $O_{\rm PM/SPT}(r)$ for (a) the original OF model at $h_{\rm zz} = 20$ and (b) the clean cluster OF model, both at the tricritical point. To reduce finite-size effects, the string length is fixed to $r = L/4$ and centered in the chain. The insets show estimates of $\eta_{\rm PM/SPT}^{\pm}(L)$ obtained from power-law fits using system sizes $L$ and $L\pm64,L\pm128$. Simulations are performed under OBC. As OBCs produce a twofold ground-state degeneracy, the string operators are measured in both global spin-flip $\mathbb{Z}_2$ even $(+)$ and odd $(-)$ sectors, which give consistent results.}
  \label{fig:string-diagnostics}
\end{figure*}

The above results suggest that the two realizations of the spontaneously fixed boundary condition are physically distinct, despite sharing the same boundary operator content.
As in the symmetry-enriched Ising CFT~\cite{Verresen2021PRX}, this distinction can be diagnosed by the $\mathbb{Z}_2^T$ charge of the disorder field.

In the TCI CFT, the lowest disorder field $\mu$ has scaling dimension $\Delta_\mu = 3/40$~\cite{zou2020prb}.
The lattice representatives in the original and cluster OF models are given in Eqs.~\eqref{eq:of-disorder-field} and~\eqref{eq:cluster-disorder-field}, respectively.
Under the antiunitary symmetry, $K \mu K = \mu$ but $K \tilde\mu K = - \tilde\mu$, so the two disorder operators carry opposite $\mathbb{Z}_2^T$ charges.
Their two-point functions define the string operators $O_{\rm PM}(r) = \langle \mu_i \mu_{i+r} \rangle$ and $O_{\rm SPT}(r) = \langle \tilde\mu_i \tilde\mu_{i+r} \rangle$.

We evaluate both strings at the spontaneously fixed boundary condition: the original OF model at $h_{\rm zz} = 20$ with the $ZZ$-type boundary deformation and the clean cluster OF model.
At the ordinary TCI QCP [Fig.~\ref{fig:string-diagnostics}(a)], $O_{\rm PM}(r)$ decays more slowly, with $\eta_{\rm PM}^{\pm} \approx 0.152$, consistent with $2 \Delta_\mu = 3/20$.
At the TCI* QCP [Fig.~\ref{fig:string-diagnostics}(b)], the slowly decaying string is instead $O_{\rm SPT}(r)$, with $\eta_{\rm SPT}^{\pm} \approx 0.15$, again matching $3/20$.
In both cases, the faster-decaying string has exponent about $2.15$, consistent within numerical resolution with the temporal descendant $\partial_\tau \mu$, whose scaling gives $2 \Delta_{\partial_\tau \mu} = 2 (1 + \Delta_\mu) = 43/20$.
Since $\mathbb{Z}_2^T$ is antiunitary and $\partial_\tau$ is time-reversal odd, $\partial_\tau \mu$ carries the opposite $\mathbb{Z}_2^T$ charge to $\mu$.

These results show that the two realizations of the spontaneously fixed boundary condition belong to different $\mathbb{Z}_2^T$ symmetry sectors, distinguished by the charge of the lowest disorder field.
This symmetry-sector distinction provides a topological invariant beyond the conventional BCFT data~\cite{Verresen2021PRX}.

\section{Extended TCI* Critical Line}
\label{sec:extended-tci-line}

Finally, we extend the TCI* realization of the cluster OF model into a continuous critical line by introducing an additional term that preserves the same $\mathbb{Z}_2 \times \mathbb{Z}_2^T$ symmetry and self-duality.
The resulting two-parameter Hamiltonian is
\begin{align}
  H = {} & - \alpha J \sum_{i} \left( Z_{i} Z_{i+1} + Z_{i} X_{i+1} Z_{i+2} \right) \nonumber \\
  & - (1 - \alpha) J \sum_{i} \left( - X_{i} + Z_{i} X_{i+1} X_{i+2} Z_{i+3} \right) \nonumber \\
  & + g \sum_{i} \left( Z_{i} X_{i+1} Z_{i+3} + Z_{i} X_{i+2} Z_{i+3} \right) \, .
  \label{eq:line_H}
\end{align}
At $\alpha = 1$ and $J = 1$, this Hamiltonian reduces to the cluster OF model studied above.
It is therefore a natural two-parameter extension of the TCI* point.
To make its self-duality explicit, we introduce Majorana operators
\begin{equation}
  \gamma_{2i-1} = \Big[ \prod_{j < i} X_{j} \Big] Z_{i} \quad \text{and} \quad \gamma_{2i} = \Big[ \prod_{j < i} X_{j} \Big] Y_{i} \, .
\end{equation}
In this representation, the Hamiltonian becomes
\begin{equation}
  \begin{aligned}
    H = {} & - {\rm i} \alpha J \sum_{i} \left( \gamma_{2i} \gamma_{2i+1} + \gamma_{2i} \gamma_{2i+3} \right) \\
    & - {\rm i} (1-\alpha) J \sum_{i} \left( \gamma_{2i} \gamma_{2i-1} + \gamma_{2i} \gamma_{2i+5} \right) \\
    & - g \sum_{i} \left( \gamma_{2i} \gamma_{2i+3} \gamma_{2i+4} \gamma_{2i+5} \right) \\
    & - g \sum_{i} \left( \gamma_{2i} \gamma_{2i+1} \gamma_{2i+2} \gamma_{2i+5} \right) \, .
  \end{aligned}
  \label{eq:line_majorana_H}
\end{equation}
It is invariant under the dual transformation $\mathcal{D}$,
\begin{equation}
  \mathcal{D}(\gamma_{r}) = \begin{cases}
      \gamma_{-r + 4} & \text{if $r$ is odd} \\
      \gamma_{-r} & \text{if $r$ is even}
    \end{cases} \, .
  \label{eq:line_duality}
\end{equation}

\begin{figure*}[t]
  \centering
  \includegraphics[width=1.0\textwidth]{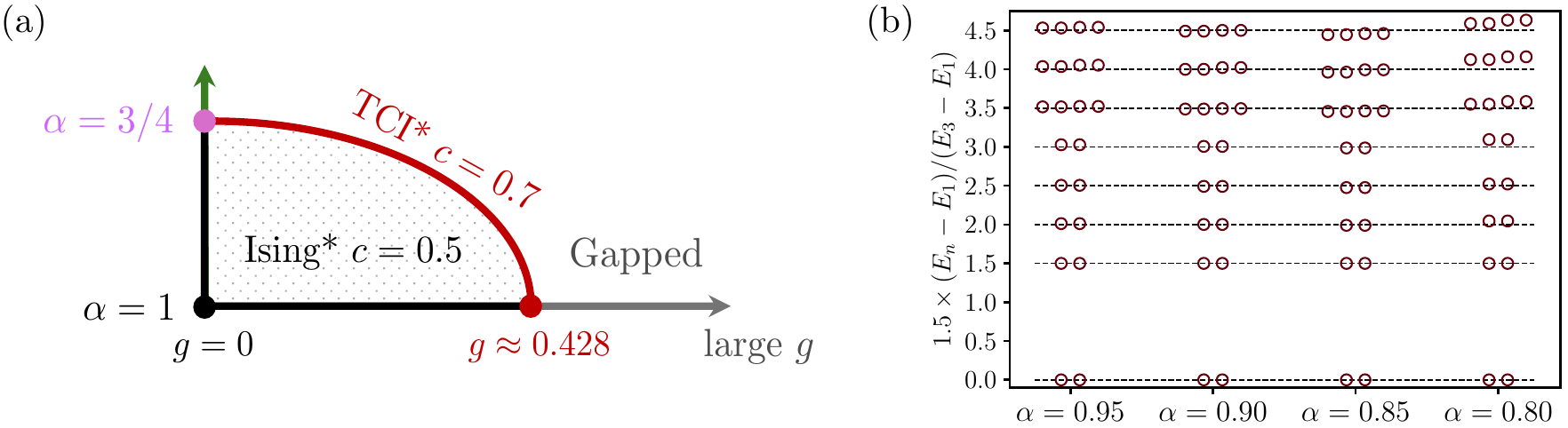}
  \caption{(a) Schematic ground-state phase diagram of the model~\eqref{eq:line_H} in the $(\alpha, g)$ plane with fixed $J = 1.0$ as the energy unit. On the $g = 0$ axis, a multicritical Lifshitz point at $\alpha = 3/4$ (characterized by a dynamical critical exponent $z = 3$) separates an Ising* CFT with central charge $c = 1/2$ from a $c = 3/2$ CFT line. For finite $g$, the Ising* criticality expands into a stable extended region, bounded by a continuous critical line belonging to the symmetry-enriched tricritical Ising (TCI*, $c = 0.7$) universality. (b) Rescaled low-lying energy spectra $1.5 \times (E_{n} - E_{1}) / (E_{3} - E_{1})$ obtained on an OBC chain of length $L = 128$. The spectra are evaluated at four representative points along the TCI* line, i.e., $(\alpha, g) = (0.95, 0.3640)$, $(0.90, 0.2975)$, $(0.85, 0.2270)$, and $(0.80, 0.1485)$; the precise locations of TCI* points are determined in Fig.~\ref{fig:line_charge}. The numerical data exhibit excellent agreement with the BCFT operator content of the spontaneous fixed boundary condition. It is noted that for $\alpha = 0.80$, its proximity to the Lifshitz point induces stronger finite-size effects; accessing larger system sizes and finer tuning of $g$ in Fig.~\ref{fig:line_charge} could further improve the data quality.}
  \label{fig:line_content}
\end{figure*}

The $g = 0$ axis provides an analytic starting point.
In this limit, Eq.~\eqref{eq:line_majorana_H} is a free Majorana fermion model~\cite{ruben2017prb}.
Fourier transforming according to
\begin{equation}
  \begin{aligned}
    \gamma_{2j-1} & = \frac{1}{\sqrt{L}} \sum_{k} {\rm e}^{-{\rm i}kj} \tilde{\gamma}_{A,k} \, , \\
    \gamma_{2j} & = \frac{1}{\sqrt{L}} \sum_{k} {\rm e}^{-{\rm i}kj} \tilde{\gamma}_{B,k} \, ,
  \end{aligned}
\end{equation}
gives
\begin{equation}
  H_{g=0} = \frac{1}{2} \sum_{k} \Psi_{k}^{\dagger} \begin{pmatrix} 
    0 & {\rm i} f(k) \\
    -{\rm i} f^{*}(k) & 0 
    \end{pmatrix} \Psi_{k} \, ,
  \label{eq:line_fourier_H}
\end{equation}
where $\Psi_{k} = (\tilde{\gamma}_{A,k}, \tilde{\gamma}_{B,k})^{\rm T}$ and
\begin{equation}
  f(k) = \alpha J \left( {\rm e}^{{\rm i}k} + {\rm e}^{2{\rm i}k} \right) + (1-\alpha) J \left( 1 + {\rm e}^{3{\rm i}k} \right) \, .
\end{equation}
The spectrum is $E(k) = \pm|f(k)|$.
Writing $w = {\rm e}^{{\rm i}k}$, the condition for zeros is determined by
\begin{equation}
  f(w) = J (w + 1) \left[ (1-\alpha) w^{2} + (2\alpha-1) w + (1-\alpha) \right] \, .
  \label{eq:line_zero_condition}
\end{equation}
There is always a root at $w = -1$, corresponding to $k = \pi$.
The remaining roots obey
\begin{equation}
  (2 - 2 \alpha) \cos{k} = 1 - 2 \alpha \, .
  \label{eq:line_additional_zeros}
\end{equation}
For $3/4 < \alpha \leq 1$, the equation has no additional roots on the unit circle, leaving a single Dirac cone at $k = \pi$ and an Ising* CFT with $c = 1/2$ in the spin model.
At $\alpha = 3/4$, the root at $w = -1$ becomes threefold and gives a Lifshitz point with dynamical exponent $z = 3$.
For $0 \leq \alpha < 3/4$, the two additional roots lie on the unit circle, so three Dirac cones give $c = 3/2$.
These three regimes on the $g = 0$ axis are summarized in Fig.~\ref{fig:line_content}(a).

\begin{figure*}[t]
  \centering
  \includegraphics[width=\textwidth]{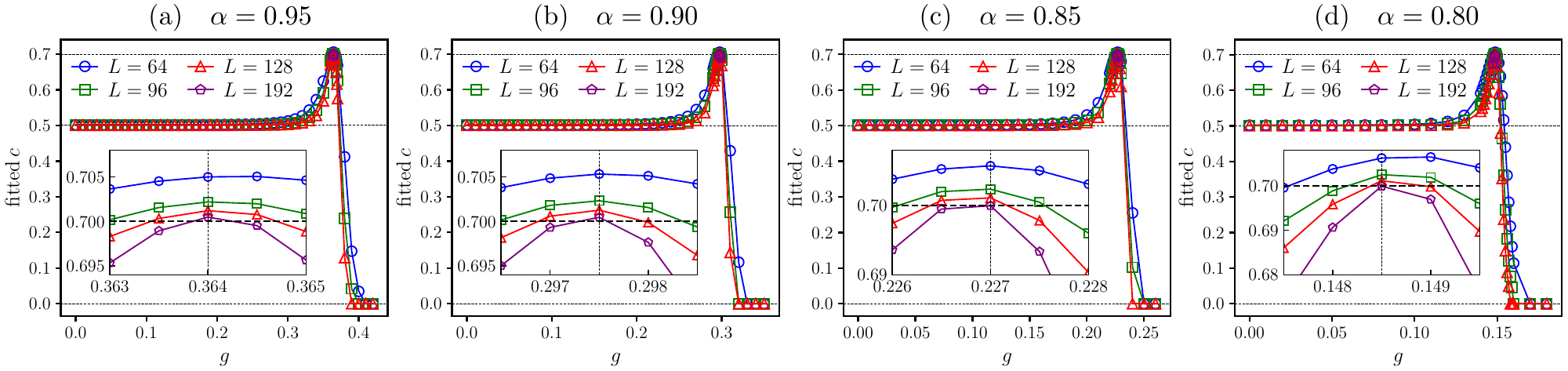}
  \caption{Extracted central charge $c$ as a function of $g$ for the model Hamiltonian~\eqref{eq:line_H} along (a) $\alpha = 0.95$, (b) $\alpha = 0.90$, (c) $\alpha = 0.85$, and (d) $\alpha = 0.80$ lines with fixed $J = 1.0$ as the energy unit. The insets highlight the finite-size effect near $\text{TCI}^*$ points, demonstrating the convergence to $c = 0.7$ at $(\alpha, g) = (0.95, 0.3640)$, $(0.90, 0.2975)$, $(0.85, 0.2270)$, and $(0.80, 0.1485)$, respectively, as $L$ increases. The extraction of the central charge is based on the bipartite von Neumann entropy; see Appendix~\ref{app:bulk-susy} for details. Simulations are performed under PBC.}
  \label{fig:line_charge}
\end{figure*}

Restoring the four-fermion interaction at $g > 0$ gives the interacting model.
Since this interaction is irrelevant with respect to the Ising* CFT, the Ising* critical regime is expected to remain stable for sufficiently small $g$.
We therefore search for TCI* criticality at the boundary of this region.
Figure~\ref{fig:line_charge} shows PBC central-charge scans along $\alpha = 0.95$, $0.90$, $0.85$, and $0.80$\,.
Each scan displays a finite Ising* regime with $c \approx 0.5$ and, on increasing $g$, a candidate TCI* point between this regime and the $c \approx 0$ region.
Near the four representative points $(\alpha,g) = (0.95,0.3640)$, $(0.90,0.2975)$, $(0.85,0.2270)$, and $(0.80,0.1485)$, the $c \approx 0.7$ feature becomes more clearly resolved with increasing system size between the $c \approx 0.5$ and $c \approx 0$ regimes.

We further examine the OBC low-energy spectra at these four points as shown in Fig.~\ref{fig:line_content}(b).
Within numerical resolution, their boundary operator contents are consistent with the spontaneously fixed boundary condition identified for the cluster OF model.
The $\alpha = 0.80$ cut has stronger finite-size effects, as it lies closest to the $\alpha = 0.75$ Lifshitz point, with visibly larger deviations from the theoretical operator-content values in the rescaled spectrum.
Together, the PBC central-charge scans and the OBC spectra support an extended TCI* critical line, rather than an isolated microscopic realization.
With the $g = 0$ analysis above, these results give the two-parameter phase diagram sketched in Fig.~\ref{fig:line_content}(a).
In this work, we restrict our attention to the $\alpha \geq 3/4$ region, which contains the Ising* regime and its TCI* boundary.

\section{Conclusion}
\label{sec:conclusion}

To summarize, we have uncovered a symmetry-enriched topological TCI realization at the tricritical point of the cluster OF model.
Using boundary magnetization, boundary-bulk correlations, and low-energy open-chain spectra, we identify the natural cluster termination with a spontaneously fixed boundary condition that hosts a twofold topological degeneracy protected by the $\mathbb{Z}_2 \times \mathbb{Z}_2^T$ symmetry.
This contrasts with the free boundary condition realized by the natural termination of the original OF model, even though the two models share the same bulk TCI CFT.

We further show that the spontaneously fixed boundary condition has two physically distinct realizations: one arises from symmetry enrichment in the cluster OF model, while the other is reached through boundary RG flow in the original OF model.
Although these realizations share the same conventional BCFT data, such as the boundary operator content and boundary $g$-function, they are distinguished by the $\mathbb{Z}_2^T$ charge of the disorder field.
Along the representative symmetry-preserving $ZZ$ and $X$ boundary deformations studied here, the twofold boundary degeneracy of the cluster model persists from weak deformation strength to the strong-deformation limit.
This weak-to-strong deformation stability cannot be inferred from the perturbative RG stability of the spontaneously fixed boundary condition alone, which only constrains sufficiently weak symmetry-preserving perturbations.
Establishing stability under arbitrary symmetry-preserving boundary deformations remains an open problem.

For the tested symmetry-preserving boundary perturbations that lift the exact degeneracy, the finite-size energy splitting of the open-chain twofold topological degeneracy is numerically consistent with exponential decay, in sharp contrast to the algebraic splitting at the Ising* critical point.
Its complete analytic origin remains open.
Finally, the TCI* realization extends along a continuous critical line in the two-parameter phase diagram.

These results show how symmetry enrichment can distinguish QCPs with identical bulk CFT through nonlocal string operators and boundary physics.
More broadly, our results may also be of interest to the general study of quantum criticality, since the symmetry-enrichment framework is not restricted in principle to conformally invariant critical points.
They also motivate further studies of critical-edge splitting in purely gapless SPT phases~\cite{Li2025SciPost} and may be relevant to quantum platforms that implement multispin interactions~\cite{Petiziol2021PRL,Zhang2022Nature}.

\begin{acknowledgments}
We thank Chengshu Li and Junchen Rong for helpful discussions.
Numerical simulations were carried out with the ITENSOR \verb|C++| package~\cite{itensor}.
X.-J. Y. was supported by the National Natural Science Foundation of China (Grant No.12405034) and a start-up grant from Eastern Institute of Technology, Ningbo.
This work is also supported by MOST 2022YFA1402701.
The work of S.Y. is supported by China Postdoctoral Science Foundation (Certificate Number: 2024M752760).
\end{acknowledgments}

\appendix
\onecolumngrid

\section{Details of DMRG simulations}
\label{app:numerical-methodology}

To calculate the quantities investigated in our work, we perform two-site DMRG simulations with matrix product states (MPSs)~\cite{White1992PRL,SCHOLLWOCK201196,Vidal2003PRL,Vidal2004PRL}. 
Throughout this work, the MPS bond dimension is maintained at a sufficiently large value such that the truncation error---defined as the sum of discarded weights during the singular value decomposition---is consistently kept below $10^{-10}$. 
Furthermore, the variational optimization process is considered converged only when the relative change in the energy between two consecutive sweeps falls below $10^{-8}$. 
These convergence criteria and strict truncation limits ensure that the obtained results are well-converged and minimize potential numerical artifacts.
Low-lying excited states are obtained sequentially by augmenting the Hamiltonian with positive penalty terms proportional to the projectors onto all previously converged lower-energy states~\cite{Stoudenmire2012ARCMP}.
For calculations with a twofold-degenerate ground state, we apply a weak boundary pinning field to select a definite boundary-polarized state;
any different treatments are specified in the corresponding figure captions (e.g., Fig.~\ref{fig:string-diagnostics}) or descriptions.
Unless otherwise stated, all numerical calculations in this work are performed using DMRG, except for the finite-size energy-splitting calculations in Sec.~\ref{sec:finite-size-splitting} and Appendix~\ref{app:splitting-analysis}, which are performed using exact diagonalization.

\section{TCI conformal field theory}
\label{app:bcft}

In this appendix, we review the conformal data of the TCI CFT in more detail than the main text, including its Virasoro algebra and emergent superconformal structures~\cite{francesco2012conformal,cardy2008boundaryconformalfieldtheory}.
TCI is the second unitary minimal model, $\mathcal{M}(4,5)$, with central charge $c = \frac{7}{10}$.
It contains six primary fields,
\begin{equation}
    \phi_{(1,1)} = I \, ,  \quad
    \phi_{(1,2)} = \epsilon \, , \quad
    \phi_{(1,3)} = \epsilon' \, , \quad
    \phi_{(1,4)} = \epsilon'' \, , \quad
    \phi_{(2,2)} = \sigma \, , \quad
    \phi_{(2,4)} = \sigma' \, ,
\end{equation}
subject to the identification $\phi_{(r,s)} = \phi_{(4-r,5-s)}$.
Their conformal weights are
\begin{equation}
    h_I = 0 \, , \qquad
    h_\epsilon = \frac{1}{10} \, , \qquad
    h_{\epsilon'} = \frac{3}{5} \, , \qquad
    h_{\epsilon''} = \frac{3}{2} \, , \qquad
    h_\sigma = \frac{3}{80} \, , \qquad
    h_{\sigma'} = \frac{7}{16} \, .
\end{equation}
For the diagonal TCI minimal model, the bulk primary fields satisfy
$h_\phi = \bar h_\phi$, so that their scaling dimensions are $\Delta_\phi = 2 h_\phi$.

The discussion above focuses on diagonal bulk primary fields and does not explicitly include the fermionic fields $\psi$ and $\bar\psi$, which carry nonzero conformal spin.
In the TCI CFT, these fields have conformal weights
\begin{equation}
  (h,\bar h)_\psi = \left( \frac{3}{5}, \frac{1}{10} \right) \, , \qquad
  (h,\bar h)_{\bar\psi} = \left( \frac{1}{10}, \frac{3}{5} \right) \, ,
\end{equation}
and therefore conformal spins
\begin{equation}
  s_\psi = h - \bar h = \frac{1}{2} \, , \qquad
  s_{\bar\psi} = h - \bar h = - \frac{1}{2} \, .
\end{equation}

The TCI CFT possesses an emergent $\mathcal{N} = 1$ superconformal symmetry, so the Virasoro algebra is enlarged to the super-Virasoro algebra.
Besides the usual Virasoro generators $L_n$, there are fermionic generators $G_r$, satisfying
\begin{equation}
  \begin{split}
    [L_m,L_n] & = (m-n) L_{m+n} + \frac{c}{12} (m^3 - m) \delta_{m+n,0} \, , \\
    \{G_r,G_s\} & = 2 L_{r+s} + \frac{c}{3} \left( r^2 - \frac{1}{4} \right) \delta_{r+s,0} \, , \\
    [L_m,G_r] & = \left( \frac{m}{2} - r \right) G_{m+r} \, .
  \end{split}
\end{equation}
Here $r$ is half-integer in the Neveu--Schwarz sector and integer in the Ramond sector.

In the Neveu--Schwarz sector, the chiral super-Virasoro mode $G_{-1/2}$ raises the left conformal weight $h$ by $\frac{1}{2}$, while the anti-chiral mode $\bar G_{-1/2}$ raises the right conformal weight $\bar h$ by $\frac{1}{2}$.
Starting from the bosonic field $\epsilon$ with $(h,\bar h)_\epsilon = (\frac{1}{10},\frac{1}{10})$, one obtains the fermionic components $G_{-1/2} \epsilon \sim \psi , \ \bar G_{-1/2} \epsilon \sim \bar\psi$ with $(h,\bar h)_\psi = (\frac{3}{5},\frac{1}{10}) , \ (h,\bar h)_{\bar\psi} = (\frac{1}{10},\frac{3}{5})$.
Acting with both modes gives the bosonic top component $G_{-1/2} \bar G_{-1/2} \epsilon \sim \epsilon'$, with $(h,\bar h)_{\epsilon'} = \left( \frac{3}{5},\frac{3}{5} \right)$.

Thus, $\epsilon$, $\psi$, $\bar\psi$, and $\epsilon'$ belong to the same bulk supermultiplet: $\psi$ and $\bar\psi$ are the fermionic superpartners of $\epsilon$, while $\epsilon'$ is the bosonic top component.
In terms of bulk scaling dimensions,
\begin{equation}
  \Delta_\epsilon = \frac{1}{5} \, , \qquad
  \Delta_\psi = \Delta_{\bar\psi} = \frac{7}{10} \, , \qquad
  \Delta_{\epsilon'} = \frac{6}{5} \, .
\end{equation}
Therefore,
\begin{equation}
  \label{eq:supersymmetry-relation}
  \Delta_\psi - \Delta_\epsilon = \Delta_{\bar\psi} - \Delta_\epsilon = \frac{1}{2} \, , \qquad
  \Delta_{\epsilon'} - \Delta_\epsilon = 1 \, .
\end{equation}
Equivalently, in the chiral or boundary language, one has
\begin{equation}
    h_{\epsilon'} - h_\epsilon = \frac{1}{2} \, ,
\end{equation}
because a single chiral super-Virasoro generator raises the conformal weight by $\frac{1}{2}$.

\section{Numerical checks of the cluster-OF phase diagram and emergent superconformal symmetry}
\label{app:bulk-susy}

\begin{figure}[t]
  \centering
  \includegraphics[width=0.9\linewidth]{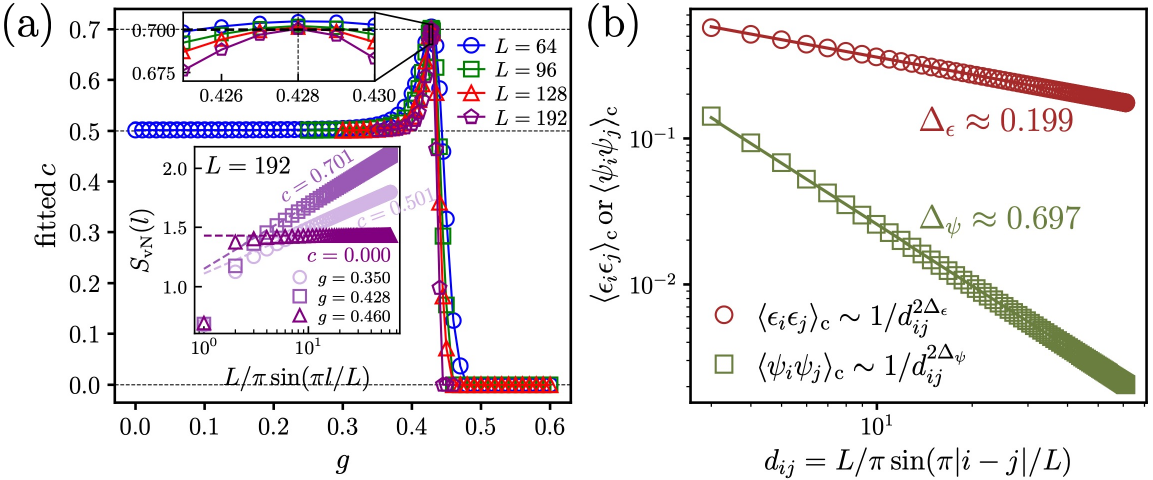}
  \caption{(a) Extracted central charge $c$ as a function of $g$ for various sizes $L$. Upper inset: finite-size effect near the TCI* QCP, demonstrating the convergence to $c = 0.7$ at $g_c = 0.428$ as $L$ increases. Lower inset: the bipartite von Neumann entanglement entropy $S_\text{vN}(l)$ versus the conformal chord length $L/\pi \sin(\pi l/L)$ for three representative points: $g = 0.350$ (on the Ising* line), $g =g_{c}$ (at the TCI* QCP), and $g = 0.460$ (on the gapped line). (b) Algebraic decay of the connected correlation functions for the bosonic field $\langle \epsilon_{i} \epsilon_{j} \rangle_\text{c}$ and the fermionic field $\langle \psi_{i} \psi_{j} \rangle_\text{c}$ at the TCI* QCP with $L = 192$. The solid lines indicate power-law fits, yielding scaling dimensions $\Delta_{\epsilon} \approx 0.199$ and $\Delta_{\psi} = 0.697$\,. All simulations are performed under PBC.}
  \label{fig:susy}
\end{figure}

This appendix provides numerical checks of the bulk ground-state phase diagram (see Fig.~\ref{fig:cluster-phase-diagram}) and the emergent supersymmetry of the cluster OF chain at the tricritical point.
We extract the central charge from the PBC entanglement scaling form of the bipartite von Neumann entropy~\cite{Pasquale_Calabrese_2004},
\begin{equation}
  S_{\rm vN}(l) \sim \frac{c}{3} \log\left[ \frac{L}{\pi} \sin\left(\frac{\pi l}{L}\right) \right] + s_0 \, .
\end{equation}
Here, $l$ is the length of a contiguous interval, $\rho_l$ is its reduced density matrix, and $S_{\rm vN}(l) = - \mathrm{Tr} \, [\rho_l \log\rho_l]$ is the von Neumann entanglement entropy.
As shown in Fig.~\ref{fig:susy}(a), the fitted central charge is $c \simeq 0.5$ for $g < g_c$, while an increasingly sharp $c \simeq 0.7$ feature develops near $g_c \simeq 0.428$ as $L$ increases; for $g>g_c$, it becomes consistent with $c = 0$.
The sharpening of this feature provides numerical evidence for a TCI point separating the $c \simeq 0.5$ and $c \simeq 0$ regimes.

We further verify the emergent supersymmetry from the connected correlation functions shown in Fig.~\ref{fig:susy}(b).
Power-law fits yield the scaling dimensions $\Delta_\epsilon \simeq 0.199$ and $\Delta_\psi \simeq 0.697$ at $g_c$, consistent with the theoretical values and the expected relation $\Delta_\psi - \Delta_\epsilon = 1/2$; see Eq.~\eqref{eq:supersymmetry-relation}.
The cluster-OF lattice representatives of the bosonic $\epsilon$ and fermionic $\psi$ fields are given in Eq.~\eqref{eq:cluster-lattice-fields};
they define the connected correlations used in Fig.~\ref{fig:susy}(b).
Here, we have symmetrized the lattice realization of the $\epsilon$ field to eliminate any ambiguity in the definition of the operator position, particularly for multi-site interactions. 
It is also worth noting that the explicit lattice representation of $\psi$ has a finite overlap with the subleading field $\bar{T}_{F}$~\cite{zou2020prb}.
Since $\Delta_{\bar{T}_{F}} = 3/2$ is larger than $\Delta_{\psi} = 7/10$, its contribution to the correlation function decays faster at large distances and is therefore asymptotically subleading.
The agreement of the fitted exponent with $\Delta_{\psi}=7/10$ indicates that this contribution does not dominate over the distances accessible here.

\section{BOE analysis of boundary scaling dimensions in the critical transverse-field Ising and cluster Ising chains}
\label{app:ising-boe}

In this appendix, we briefly review the symmetry-enriched Ising criticality and use the BOE analysis to understand the boundary scaling dimensions of the spin and energy fields at the Ising and Ising* QCPs.
The critical transverse-field Ising and cluster Ising chains realize the conventional Ising and symmetry-enriched Ising (Ising*) critical points, respectively.
Their Hamiltonians are
\begin{equation}
  \begin{aligned}
    H_{\rm TFI} & = - \sum_i \left( Z_i Z_{i+1} + X_i \right) \, ,\\
    H_{\rm CI} & = - \sum_i \left( Z_i Z_{i+1} + Z_{i-1} X_i Z_{i+1} \right) \, .
  \end{aligned}
  \label{eq:ising-hamiltonians}
\end{equation}
Both models preserve the global spin-flip $\mathbb{Z}_{2}$ symmetry and spinless time-reversal symmetry $\mathbb{Z}_{2}^{T}$, and their critical points are described by the $(1+1)$-dimensional Ising CFT.
For Ising CFT~\cite{francesco2012conformal,cardy2008boundaryconformalfieldtheory}, its three diagonal primary fields are $I$, $\sigma$, and $\epsilon$, with conformal weights $(h,\bar h)=(0,0)$, $(1/16,1/16)$, and $(1/2,1/2)$, and scaling dimensions $\Delta = 0$, $1/8$, and $1$, respectively.
Their fusion rules are
\begin{equation}
  \sigma \times \sigma = I + \epsilon \, , \qquad
  \sigma \times \epsilon = \sigma \, , \qquad
  \epsilon \times \epsilon = I \, .
\end{equation}
The Cardy states $|I\rangle$ and $|\epsilon\rangle$ correspond to the two fixed boundary conditions, while $|\sigma\rangle$ corresponds to the free boundary condition;
accordingly, the spontaneously fixed boundary condition is $|\!+\!\&-\rangle = |I\rangle + |\epsilon\rangle$.
The nonlocal disorder field $\mu$ has scaling dimension $\Delta_\mu = 1/8$~\cite{francesco2012conformal}.
The disorder-string representatives in the two Ising CFT realizations are, respectively, given in Eqs.~\eqref{eq:of-disorder-field} and~\eqref{eq:cluster-disorder-field};
their $\mathbb{Z}_{2}^{T}$ charges are opposite: $K \mu K = \mu$ and $K \tilde{\mu} K = - \tilde{\mu}$.
Thus, the transverse-field Ising and cluster Ising chains realize symmetry-distinct critical points despite sharing the same bulk Ising CFT.
Under natural OBC, the transverse-field Ising chain realizes the free boundary condition, whereas the cluster Ising chain realizes the spontaneously fixed boundary condition.
The bulk-boundary correlation calculations of Ref.~\cite{Yu2022PRL} give the boundary scaling dimensions $\Delta_\sigma^{\rm s} = 1/2$ and $\Delta_\epsilon^{\rm s} = 2$ for the transverse-field Ising chain, whereas for the cluster Ising chain, the boundary scaling dimensions are $\Delta_\sigma^{\rm s} = \Delta_\epsilon^{\rm s} = 2$.
These boundary dimensions can be understood from the same BOE reasoning used in Sec.~\ref{sec:boundary-correlations-boe}.

We first consider the bulk spin field, whose fusion rule is $[\sigma] \times [\sigma] = [I] + [\epsilon]$.
For the free boundary condition $|\sigma\rangle$, the boundary spectrum is likewise $[\sigma] \times [\sigma] = [I] + [\epsilon]$, so the boundary primary $[\epsilon]$ can appear in the BOE.
The leading nontrivial boundary primary therefore has scaling dimension $\Delta_{\epsilon}^{\rm bdy} = 1/2$.
For either fixed boundary condition, $|I\rangle$ or $|\epsilon\rangle$, the boundary spectrum contains only the identity, $[I] \times [I] = [\epsilon] \times [\epsilon] = [I]$.
The leading nontrivial contribution to the BOE therefore comes from the boundary stress tensor with $\Delta_T = 2$.

For the bulk energy field, the fusion rule is $[\epsilon] \times [\epsilon] = [I]$.
Thus, for both free and fixed boundary conditions, only the identity can appear as a boundary primary in its BOE.
The leading nontrivial correction is again given by the boundary stress tensor with $\Delta_T = 2$.

\section{Character-based expectation for TCI* edge-mode splitting}
\label{app:splitting-analysis}

As shown in the main text, once $\mathbb{Z}_{2} \times \mathbb{Z}_{2}^{T}$-preserving boundary perturbations are added, the energy splitting between the two nearly degenerate ground states decays exponentially with system size.
This behavior is qualitatively different from that of the critical cluster Ising model, where the corresponding splitting is algebraic.

We first derive the possible algebraic contribution by following the argument for the cluster Ising model in Ref.~\cite{Verresen2021PRX}.
After the twist operation, the disorder operator $\mu$ transforms nontrivially under the antiunitary time-reversal symmetry $\mathbb{Z}_{2}^{T}$.
In the bulk, $\mu$ is not identical to the TCI primary field $\epsilon''$. 
However, when a disorder line terminates on a fixed boundary, its endpoint necessarily flips the sign of the boundary spin. 
Equivalently, the endpoint of the symmetry-flux line exchanges the two fixed boundary states $|+\rangle \leftrightarrow |-\rangle $.
Thus, at the boundary, $\mu(0)$ plays the role of the boundary-condition-changing operator between $|+\rangle$ and $|-\rangle$. 
For the Ising CFT, this boundary-condition-changing operator is $\epsilon$, with scaling dimension $\Delta_\epsilon^{\rm bdy} = 1/2$. 
For the TCI CFT, the analogous operator is $\epsilon''$, with scaling dimension $\Delta_{\epsilon''}^{\rm bdy} = 3/2$.

In the critical Ising case, the primary field $\epsilon$ corresponds to the character $\chi_{2,1}$.
By analogy, in the TCI case we should consider the character associated with $\epsilon''$, namely $\chi_{3,1}$. 
The corresponding irreducible character is~\cite{feverati2006renormalisation}
\begin{equation}
  \begin{split}
    q^{-h_{r,s}+c/24}\chi_{r,s}(q) \rightarrow q^{-\frac{353}{240}}\chi_{3,1}(q) = 1+q+2q^2+2q^3+3q^4+4q^5+\cdots \, .
  \end{split}
\end{equation}
The leading descendant contribution relevant for the energy splitting appears at level $n = 5$.
Therefore the corresponding boundary scaling dimension is $\Delta_5 = \Delta_{\epsilon''}^{\rm bdy} + 5 = \frac{3}{2} + 5 = \frac{13}{2}$.
Using the standard boundary perturbation estimate $\delta E(L) \sim L^{-(2\Delta_5 - 1)}$, one obtains the algebraic exponent $\beta = 2\Delta_5 - 1 = 12$.
Therefore, by direct analogy with the cluster Ising model, one might expect a power-law splitting induced by a boundary operator that preserves the full $\mathbb{Z}_2 \times \mathbb{Z}_2^T$ symmetry.

However, our numerical results in the main text show a qualitatively different behavior. 
For example, the boundary perturbation $\lambda( X_1 Z_2 Z_3 + Z_{L-2} Z_{L-1} X_L )$ with $\lambda = 0.1$ produces an exponential, rather than algebraic, decay of the ground-state energy splitting.

A possible explanation is that the ordinary Virasoro boundary character is too coarse for the symmetry-enriched problem. 
It correctly specifies the boundary operator content of the TCI CFT, but it does not keep track of the $\mathbb{Z}_2^T$ quantum numbers of these operators. 
Therefore, a boundary operator that appears in the ordinary character may still be forbidden from contributing to the edge-mode splitting by the microscopic antiunitary $\mathbb{Z}_2^T$ symmetry. 

\begin{figure}[t]
  \centering
  \includegraphics[width=0.8\linewidth]{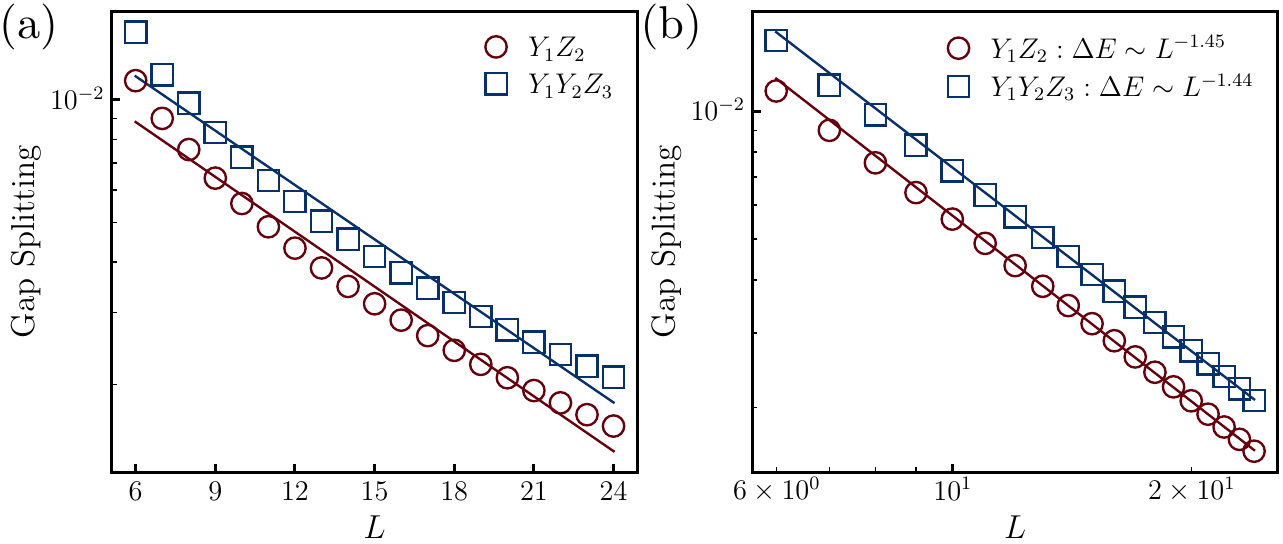}
  \caption{Algebraic splitting of the topological twofold ground-state degeneracy at the $\text{TCI}^*$ QCP. Energy splitting $\delta E(L)$ as a function of the size $L$ under boundary perturbations that explicitly break the $\mathbb{Z}_{2} \times \mathbb{Z}_{2}^{T}$ protecting symmetry. Two representative terms are considered (with $\lambda = 0.1$): $\lambda \left( Y_{1}Z_{2} + Z_{L-1}Y_{L} \right)$ [red circles] which breaks the time-reversal $\mathbb{Z}_{2}^{T}$ symmetry, and $\lambda \left( Y_{1}Y_{2}Z_{3} + Z_{L-2} Y_{L-1} Y_{L} \right)$ [blue squares] which breaks the global spin-flip $\mathbb{Z}_{2}$ symmetry. (a) Semi-log plots of the energy splitting. The solid lines represent exponential fits $\delta E(L) \sim \exp(-L/\xi)$, which clearly deviate from the numerical data. (b) The same data plotted  on a log-log scale. The straight solid lines indicate excellent fits to an algebraic decay $\delta E(L) \sim L^{-a}$. Simulations are performed under OBC using exact diagonalization.}
  \label{fig:algebraic_power_decay}
\end{figure}

This distinction is important. 
If the protecting symmetry is not imposed, we can indeed find boundary operators that lead to algebraic splitting, as shown in Fig.~\ref{fig:algebraic_power_decay}.
In contrast, for all $\mathbb{Z}_2 \times \mathbb{Z}_2^T$-preserving boundary perturbations tested in the main text, the splitting is numerically consistent with exponential decay over the accessible system sizes.
This suggests that the absence of algebraic splitting might be caused by an additional $\mathbb{Z}_2^T$ selection rule beyond the ordinary Virasoro boundary spectrum.
A complete CFT formulation of this selection rule, especially for an antiunitary symmetry, remains an interesting open problem.

\section{Boundary RG flow and boundary $g$-function diagnostics}
\label{app:boundary-rg}

This appendix contains two complementary numerical diagnostics for the boundary RG flow induced by the $ZZ$-type boundary deformation in the original OF model.
First, we locate the unstable degenerate boundary fixed point, thereby verifying the boundary phase diagram discussed in Sec.~\ref{sec:boundary-deformations}.
Second, we extract the boundary $g$-function as an independent diagnostic of the corresponding boundary conditions and RG-flow structure.

\subsection{Numerical identification of the degenerate boundary fixed point}

As discussed in the main text, introducing the boundary term, $- h_\text{zz} ( Z_{1}Z_{2} + Z_{L-1}Z_{L} )$, to the original OF model can drive a boundary phase transition from the conformal free boundary condition to the spontaneously fixed boundary condition.
This transition is mediated by an unstable degenerate boundary condition.
In this subsection, we detail the numerical determination and verification of this degenerate boundary state.

According to TCI BCFT, the operator content of the degenerate state can be obtained from the fusion rule $[\sigma] \times [\sigma] = [I] + [\epsilon] + [\epsilon'] + [\epsilon'']$.
This yields a specific set of nontrivial boundary primary fields with scaling dimensions $\Delta_{\epsilon}^\text{bdy} = 1/10$, $\Delta_{\epsilon'}^\text{bdy} = 3/5$, and $\Delta_{\epsilon''}^\text{bdy} = 3/2$~\cite{Iino2020PRB}.
This information provides a sensitive metric to locate the position of the degenerate state numerically: the gap ratio $(E_{3} - E_{1}) / (E_{2} - E_{1})$.
For the degenerate boundary state, this ratio is expected to approach $6$;
in contrast, the ratio converges to $4/3$ for the free boundary state and diverges for the spontaneously fixed boundary state due to the twofold ground-state degeneracy.

\begin{figure}[t]
  \centering
  \includegraphics[width=0.9\linewidth]{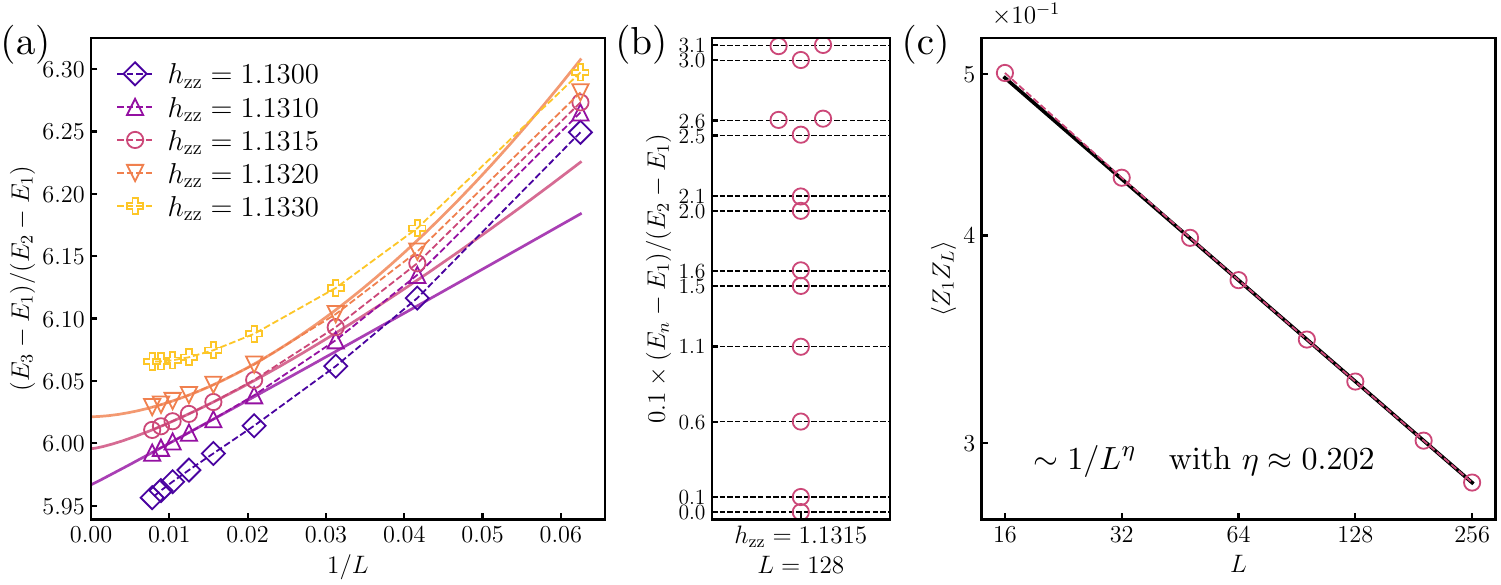}
  \caption{(a) Finite-size scaling of the energy-gap ratio $(E_{3} - E_{1})/(E_{2}-E_{1})$ versus $1/L$ for different boundary-field strengths $h_\text{zz}$. The thermodynamic limit, i.e., $1/L \to 0$, of this ratio serves as a sensitive diagnostic tool here: it is $6$ for the degenerate fixed boundary condition, $4/3$ for the free boundary condition, and diverges at the spontaneously fixed boundary condition. A power-law extrapolation finds this ratio close to $6$ near $h_\text{zz} \approx 1.1315$, locating the estimated degenerate boundary condition on the boundary RG line [see the inset of Fig.~\ref{fig:boundary-rg-comparison}(a1)]. (b) Rescaled energy spectrum at the estimated degenerate fixed point $h_\text{zz} = 1.1315$ for $L = 128$. The extracted energy levels and their corresponding degeneracies exhibit a consistent correspondence with the operator content of the underlying BCFT at the degenerate fixed point. (c) Boundary spin-spin correlation function $\langle Z_{1} Z_{L} \rangle$ as a function of system size $L$ on a logarithmic scale at $h_\text{zz} = 1.1315$\,. The data are fitted by a power-law decay $\sim 1/L^{\eta}$, where $\eta \approx 0.202$ is the anomalous exponent. This exponent is close to twice the smallest scaling dimension $\Delta_{\epsilon}^{\rm bdy} = 1/10$ within the corresponding BCFT operator content shown in (b). All simulations are performed for the original OF model with the boundary term $-h_\text{zz}(Z_{1}Z_{2} + Z_{L-1}Z_{L})$ under OBC.}
  \label{fig:degenerate_state}
\end{figure}

To pinpoint the degenerate boundary state, we compute the finite-size energy ratio for various boundary strengths.
As shown in Fig.~\ref{fig:degenerate_state}(a), by extrapolating to the thermodynamic limit, we find that the ratio is close to $6$ near $h_\text{zz} \approx 1.1315$, locating the estimated degenerate boundary condition.

To further verify that the system at $h_\text{zz} \approx 1.1315$ indeed realizes the degenerate boundary condition, we extract the low-lying energy spectrum as shown in Fig.~\ref{fig:degenerate_state}(b).
The obtained spectrum is consistent with the operator content of the degenerate boundary condition after suitable rescaling.
In addition, in Fig.~\ref{fig:degenerate_state}(c), we calculate the finite-size scaling behavior of the boundary spin-spin correlation;
the extracted exponent $\eta \approx 0.202$ is close to twice the lowest non-trivial scaling dimension, $2\Delta_{\epsilon}^\text{bdy} = 1/5$, present in the operator content.

\subsection{Boundary $g$-function from wavefunction overlaps}

To complement the spectral and correlation diagnostics above, we further extract the boundary $g$-function for the free, degenerate, and spontaneously fixed boundary conditions using wavefunction overlaps.

The boundary $g$-function, introduced by Affleck and Ludwig~\cite{affleck1991prl}, determines the universal boundary entropy $\ln g_a$ associated with a conformal boundary condition $a$.
In BCFT, it is defined as the overlap between the bulk vacuum state $|0\rangle$ and the corresponding conformal boundary state: $g_{a} = \langle 0 | a \rangle$.
Here, the boundary $g$-function should not be confused with the coupling strength $g$ in the lattice Hamiltonian~\eqref{eq:model}.
Under unitary boundary RG flow, the $g$-theorem states that $g$ decreases from ultraviolet to infrared fixed points.

Recently, wavefunction-overlap methods have been shown to extract universal conformal data, including the boundary $g$-function, from overlaps of low-energy eigenstates under different boundary conditions (or different deformations of the Hamiltonian).
In our case, we consider a system of size $L$ and construct three distinct ground states:
(i) $|\Psi_{00}\rangle$, the ground state of the uniform PBC system;
(ii) $|\Psi_{0a}\rangle$, the ground state where a boundary condition $a$ is implemented on the bond $(L/2, L/2+1)$;
and (iii) $|\Psi_{aa}\rangle$, the ground state where the same boundary condition $a$ is simultaneously realized on two bonds, i.e., $(L, 1)$ and $(L/2, L/2+1)$.
When an open-chain segment has a twofold-degenerate ground state, we use its $\mathbb{Z}_2$-even state in evaluating the wavefunction overlaps.

Following the wavefunction-overlap method proposed in Ref.~\cite{zou2024gfunc}, the boundary $g$-function associated with the conformal boundary state $|a\rangle$ can be directly estimated by the ratio of wavefunction overlaps:
\begin{equation}
  g_{a} = \frac{\langle\Psi_{00}|\Psi_{0a}\rangle}{\langle\Psi_{aa}|\Psi_{0a}\rangle} \, .
\end{equation}

\begin{figure}[t]
  \centering
  \includegraphics[width=1.0\linewidth]{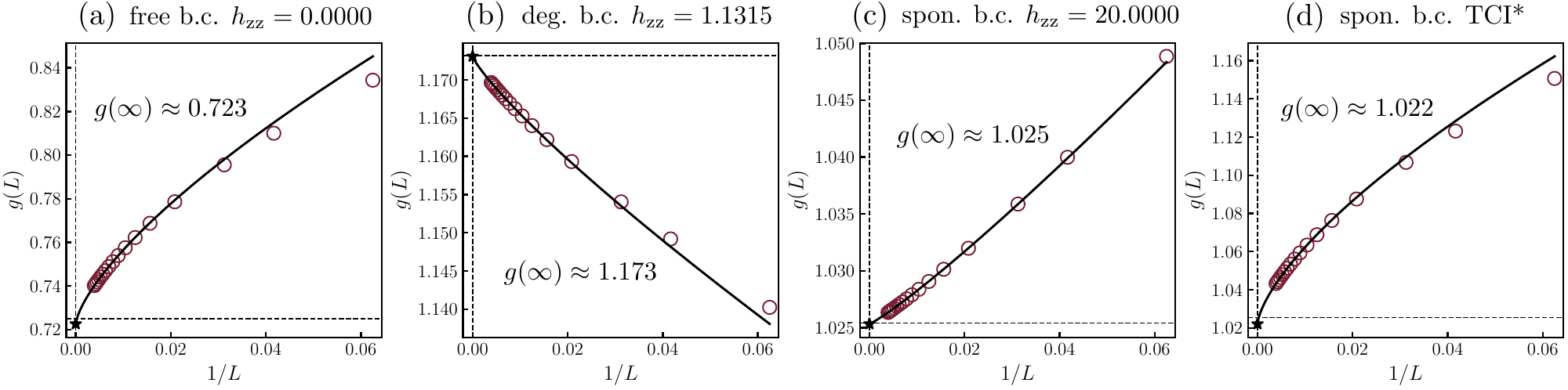}
  \caption{Finite-size scaling of the boundary $g$-function, $g(L)$, calculated via the wavefunction-overlap method as a function of the inverse system size $1/L$. Results are shown for three distinct conformal boundary conditions realized by the original OF model with the boundary term $-h_\text{zz}(Z_{1}Z_{2} + Z_{L-1}Z_{L})$: (a) the free boundary condition at $h_\text{zz} = 0$, (b) the degenerate boundary condition at $h_\text{zz} = 1.1315$, and (c) the spontaneously fixed boundary condition at $h_\text{zz} = 20$, as well as (d) the spontaneously fixed boundary condition realized by the clean cluster OF model. The solid curves represent power-law extrapolations to the thermodynamic limit, yielding $g(\infty) \approx 0.723$, $1.173$, $1.025$, and $1.022$, respectively, denoted by stars. These extrapolated values are in close agreement with the exact analytical predictions~\cite{CHIM1996IJMPA}: $g_\text{free} = \sqrt{2} C \approx 0.725$, $g_\text{deg.} = \sqrt{2} \gamma^{2} C \approx 1.173$, and $g_\text{spon.} = 2 C \approx 1.025$, as indicated by the horizontal dashed lines, where $C = \sqrt{\sin(\pi/5)/\sqrt{5}}$ and $\gamma = \sqrt{\sin(2\pi/5)/\sin(\pi/5)}$.}
  \label{fig:gfunction}
\end{figure}

Figure~\ref{fig:gfunction} shows the finite-size extrapolations of $g_{a}$ for the free, degenerate, and spontaneously fixed boundary conditions in the original OF model, as well as the naturally realized spontaneously fixed boundary condition in the cluster OF model.
The extrapolated values are $g_\text{free}\approx0.723$, $g_\text{deg.}\approx1.173$, $g_\text{spon.}\approx1.025$ for the original OF model, and $g_\text{spon.}\approx1.022$ for the cluster OF model.
They agree with the exact BCFT values~\cite{CHIM1996IJMPA}
\begin{equation}
  g_\text{free} = \sqrt{2} C \approx 0.725 \, ,\qquad
  g_\text{deg.} = \sqrt{2} \gamma^{2} C \approx 1.173 \, ,\qquad
  g_\text{spon.} = 2 C \approx 1.025 \, ,
\end{equation}
where $C = \sqrt{\sin(\pi/5)/\sqrt{5}}$ and $\gamma = \sqrt{\sin(2\pi/5)/\sin(\pi/5)}$.
In particular, the hierarchy $g_\text{deg.} > g_\text{free}$ and $g_\text{deg.} > g_\text{spon.}$ is consistent with the Affleck--Ludwig $g$-theorem and supports the boundary RG-flow structure in Fig.~\ref{fig:boundary-rg-comparison}(a1).

\section{Additional numerical results for $ZZ$-type boundary deformations in the cluster OF Model}
\label{app:boundary-perturbations}

As discussed in the main text, the twofold topological edge-mode degeneracy of the cluster OF model remains robust under the same $ZZ$-type boundary term that drives the original OF model from the free to the spontaneously fixed boundary condition.
Here, we provide extended low-energy spectra over a wider range of $h_\text{zz}$ to further support this result.

\begin{figure}[t]
  \centering
  \includegraphics[width=1.0\linewidth]{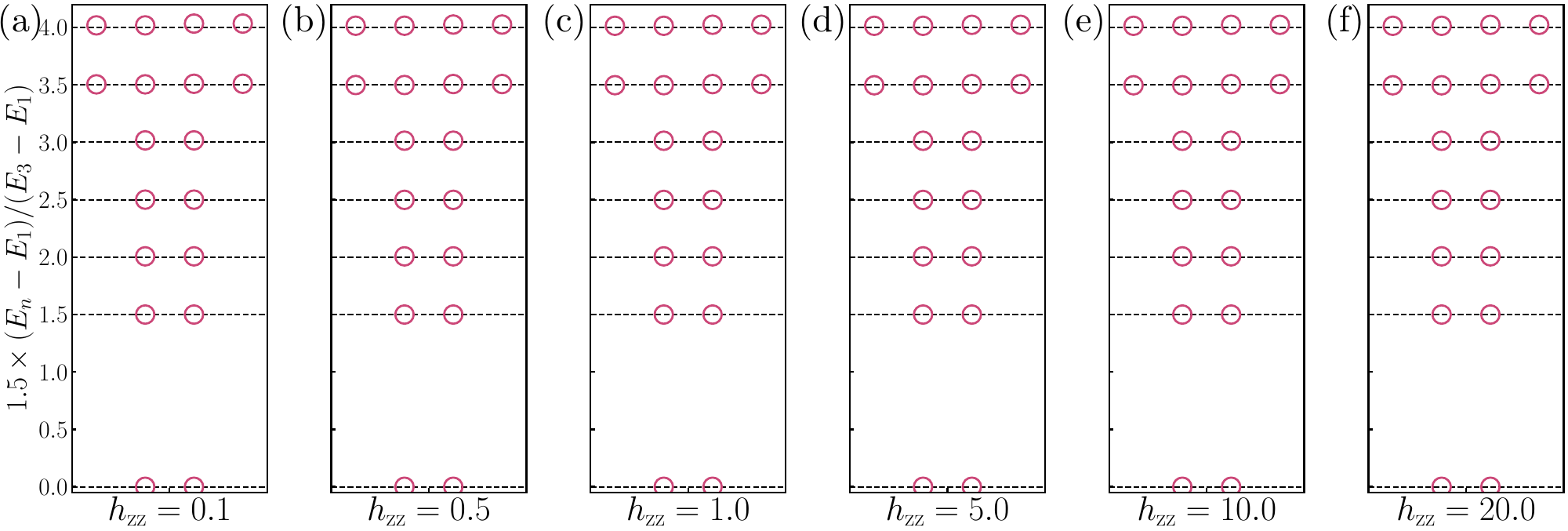}
  \caption{(a)--(f) Rescaled energy spectra, $1.5 (E_{n} - E_{1}) / (E_{3} - E_{1})$, of the cluster OF model at the TCI* QCP in the presence of a boundary term $H_\text{bdy} = - h_\text{zz} \left( Z_{1}Z_{2} + Z_{L-1}Z_{L} \right)$. The panels illustrate the spectral evolution across a wide range of finite perturbation strengths, specifically for $h_\text{zz} = 0.1$, $0.5$, $1.0$, $5.0$, $10.0$, and $20.0$, respectively, from left to right. The numerically extracted low-energy spectra (open circles) are consistent, within numerical resolution, with the BCFT operator content (dashed lines) of the spontaneously fixed boundary condition. The simulated system size is $L = 128$ under OBC.}
  \label{fig:zzpstable}
\end{figure}

\begin{figure}[t]
  \centering
  \includegraphics[width=1.0\linewidth]{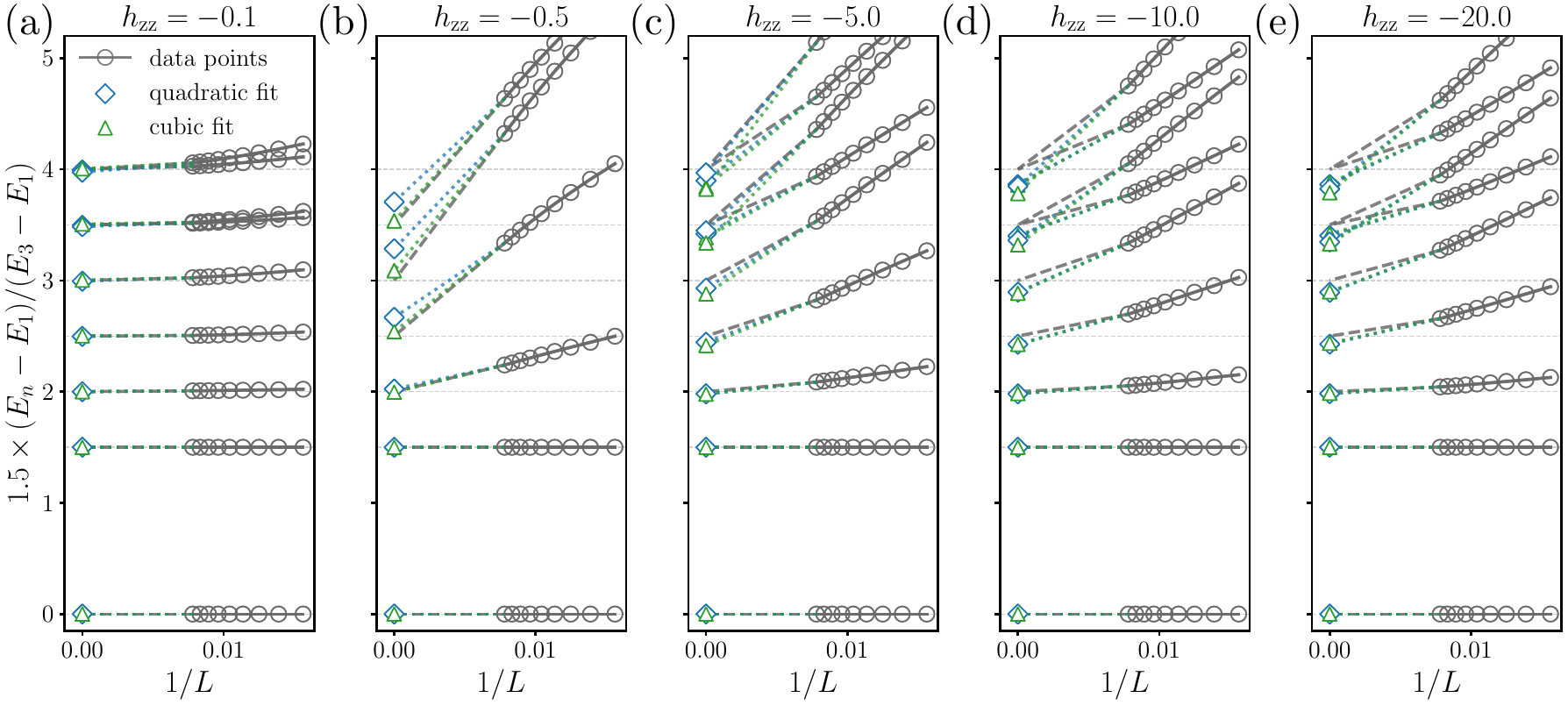}
  \caption{(a)--(e) Rescaled energy spectra, $1.5 (E_{n} - E_{1}) / (E_{3} - E_{1})$, of the cluster OF model at the TCI* QCP in the presence of a boundary term $H_\text{bdy} = - h_\text{zz} \left( Z_{1}Z_{2} + Z_{L-1}Z_{L} \right)$. The panels illustrate the spectral evolution across a wide range of finite perturbation strengths, specifically for $h_\text{zz} = -0.1$, $-0.5$, $-5.0$, $-10.0$, and $-20.0$, respectively, from left to right.
  Each marker denotes a twofold degeneracy.
  Finite-size extrapolations are applied to extract the thermodynamic limits, which closely align with the corresponding scaling dimensions (gray dashed lines as guides for the eye). Simulations are performed under OBC.}
  \label{fig:zzmstable}
\end{figure}

\begin{figure}[tb]
  \centering
  \includegraphics[width=\linewidth]{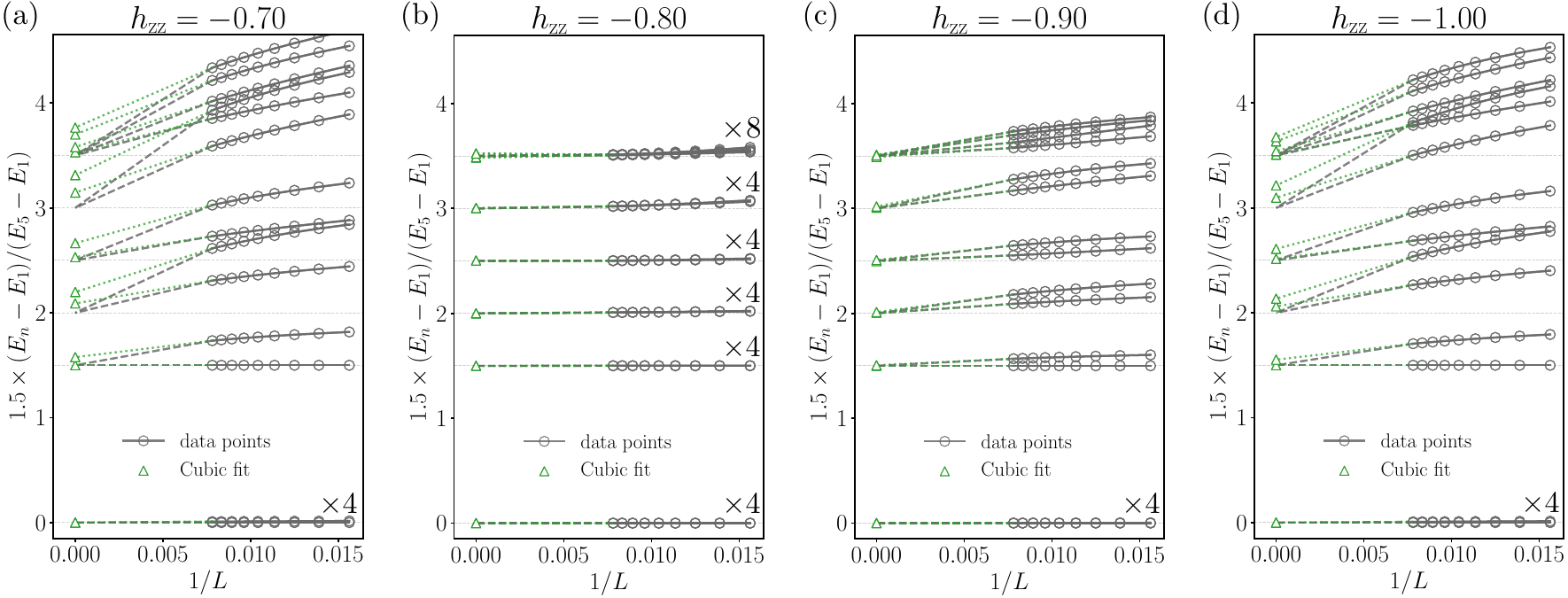}
  \caption{(a)--(d) Rescaled energy spectra, $1.5 (E_{n} - E_{1}) / (E_{3} - E_{1})$, of the cluster OF model at the TCI* QCP in the presence of a boundary term $H_\text{bdy} = - h_\text{zz} \left( Z_{1}Z_{2} + Z_{L-1}Z_{L} \right)$. The panels illustrate the spectral evolution across a narrower range of finite perturbation strengths, specifically for $h_\text{zz} = -0.7$, $-0.8$, $-0.9$, and $-1.0$, respectively, from left to right.
  Each marker denotes a twofold degeneracy (higher degeneracies are indicated by the numbers).
  Finite-size extrapolations are applied to extract the thermodynamic-limit results, which closely align with the corresponding scaling dimensions (gray dashed lines as guides for the eye). Only the first 28 lowest-lying energy levels are displayed here. Simulations are performed under OBC.}
  \label{fig:fourfold}
\end{figure}

In particular, the rescaled energy spectra $1.5 (E_{n}-E_{1}) / (E_{3}-E_{1})$ for various positive perturbation strengths, specifically $h_\text{zz} = 0.1, 0.5, 1.0, 5.0, 10.0,$ and $20.0$, are shown in Fig.~\ref{fig:zzpstable}. 
Complementarily, Fig.~\ref{fig:zzmstable} displays the corresponding results for negative perturbation strengths, covering $h_\text{zz} = -0.1, -0.5, -5.0, -10.0,$ and $-20.0$\,.
The spectra are consistent with the corresponding BCFT operator content of the spontaneously fixed boundary condition.
Notably, for $h_\text{zz} < 0$, the results exhibit significantly more pronounced finite-size effects than in the positive-$h_\text{zz}$ case.
To account for these finite-size corrections, we apply a systematic finite-size extrapolation to the thermodynamic limit for the data shown in Fig.~\ref{fig:zzmstable} and Fig.~\ref{fig:fourfold}. 
Specifically, for each energy level $E_{n}$, we fit its gap with respect to the ground-state energy $E_{1}$ using a quadratic or cubic polynomial in terms of $1/L$:
\begin{equation}
  E_{n} - E_{1} \sim \frac{a_{n}}{L} + \frac{b_{n}}{L^{2}} \quad \text{or} \quad E_{n} - E_{1} \sim \frac{a_{n}}{L} + \frac{b_{n}}{L^{2}} + \frac{c_{n}}{L^{3}} \, .
\end{equation}
After extracting the leading scaling coefficients $a_{n}$ from the fitting, the rescaled energy spectrum in the thermodynamic limit can be estimated via $\frac{3 (E_{n} - E_{1})}{2 (E_{3} - E_{1})} \approx \frac{3 a_{n}}{2 a_{3}}$. 

Interestingly, as shown in Fig.~\ref{fig:fourfold}, a striking feature emerges in the vicinity of $h_\text{zz} \approx -1.0$, where the low-energy spectrum exhibits a distinct fourfold degeneracy.
To understand this phenomenon, we consider the full Hamiltonian of the cluster OF model under the $ZZ$-type boundary deformation and conjugate it by the open-chain SPT entangler;
the resulting Hamiltonian is 
\begin{equation}
  H = H'_\text{OF} - (h_\text{zz} + 1) (Z_{1} Z_{2} + Z_{L-1} Z_{L}) + g (Z_{1}Z_{2}X_{3} + X_{L-2}Z_{L-1}Z_{L}) \, ,
\label{eq:spt_H}
\end{equation}
where $H'_\text{OF}$ denotes the original OF Hamiltonian defined on the internal sites from $2$ to $L-1$. 
Crucially, since the boundary Pauli operators $Z_{1}$ and $Z_{L}$ commute with the entire Hamiltonian, we can treat them as classical numbers, i.e., $Z_{1} = z_{1}$ and $Z_{L} = z_{L}$. 
The Hamiltonian then reduces to 
\begin{equation}
  H = H'_\text{OF} - (h_\text{zz} + 1) (z_{1} Z_{2} + z_{L} Z_{L-1}) + g (z_{1}Z_{2}X_{3} + z_{L}X_{L-2}Z_{L-1}) \, .
\end{equation}
Under the global $\mathbb{Z}_{2}$ symmetry, the boundary classical variables transform as $z_{1}\to-z_{1}$ and $z_{L}\to-z_{L}$. 
This symmetry guarantees that the energy spectrum must at least be twofold degenerate, establishing a pairwise connection between the sectors $(z_{1}, z_{L}) = (+, +)$ and $(-,-)$, as well as $(+,-)$ and $(-,+)$.

Since our boundary tuning parameter is $h_\text{zz}$, we mainly consider the effects of the boundary terms $-(h_\text{zz}+1) z_{1} Z_{2}$ and $-(h_\text{zz}+1) z_{L} Z_{L-1}$. 
As illustrated by our numerical calculations of the boundary spin scaling dimension in the main text, we have $\Delta_{\sigma}^\text{s} \approx 1.5$ for the free boundary condition. 
Within the framework of BCFT, a boundary operator with a scaling dimension greater than unity acts as an irrelevant perturbation. 
Consequently, the corresponding weak boundary perturbations are RG irrelevant, and their couplings flow to zero at the free boundary fixed point realized by $H'_\text{OF}$. 
A sufficiently strong microscopic boundary field can nevertheless drive a boundary phase transition from the free to the fixed $\pm$ boundary condition.

This picture provides a reasonable explanation for the observed behavior of the energy spectrum as a function of $h_\text{zz}$:
\begin{itemize}
  \item[i.] Near the point $h_\text{zz} = -1.0$, the boundary spin fields $-(h_\text{zz}+1) z_{1} Z_{2}$ and $-(h_\text{zz}+1) z_{L} Z_{L-1}$ either vanish identically or remain very weak. Since these fields remain within the basin of attraction of the free boundary fixed point, the system flows back to the stable free boundary condition. Without sufficiently strong boundary fields to produce an energy difference between the symmetry sectors, the degeneracy of all four sectors, i.e., $(+,+)$, $(-,-)$, $(+,-)$, and $(-,+)$, remains intact. This explains the fourfold degeneracy observed within a finite parameter region around $h_\text{zz} = -1.0$\,. Interestingly, our numerical results show that while the fourfold degeneracy exists at $h_\text{zz} = -1.0, -0.9, -0.8$, and $-0.7$, the finite-size effects are visibly smaller at $h_\text{zz} = -0.8$, indicating that the point of minimum finite-size corrections shifts slightly away from the value $h_\text{zz} = -1.0$\,. This behavior can be understood by considering the extra perturbation terms, i.e., $g z_{1}Z_{2}X_{3}$ and $g z_{L} X_{L-2}Z_{L-1}$. Although these terms are also irrelevant perturbations, their presence can affect the finite-size corrections.
  \item[ii.] As $|h_\text{zz} + 1|$ increases beyond a critical value, the system crosses the finite-coupling boundary transition and flows toward the fixed boundary condition, locking the boundary spins into ordered states. Compared with the parallel-alignment sectors, $(+,+)$ and $(-,-)$, the opposite-alignment sectors, $(+,-)$ and $(-,+)$, have a higher energy, which yields the observed twofold degeneracy. 
\end{itemize}

To verify this picture, we perform two complementary numerical analyses. 
We first use the original OF model under a boundary Zeeman field to determine the field scale at which the free boundary condition is destabilized. 
We then show that, in a simplified cluster OF model, the fourfold degeneracy is lifted roughly at the same scale, supporting the boundary RG interpretation proposed above.

\begin{figure}[t]
  \centering
  \includegraphics[width=0.8\linewidth]{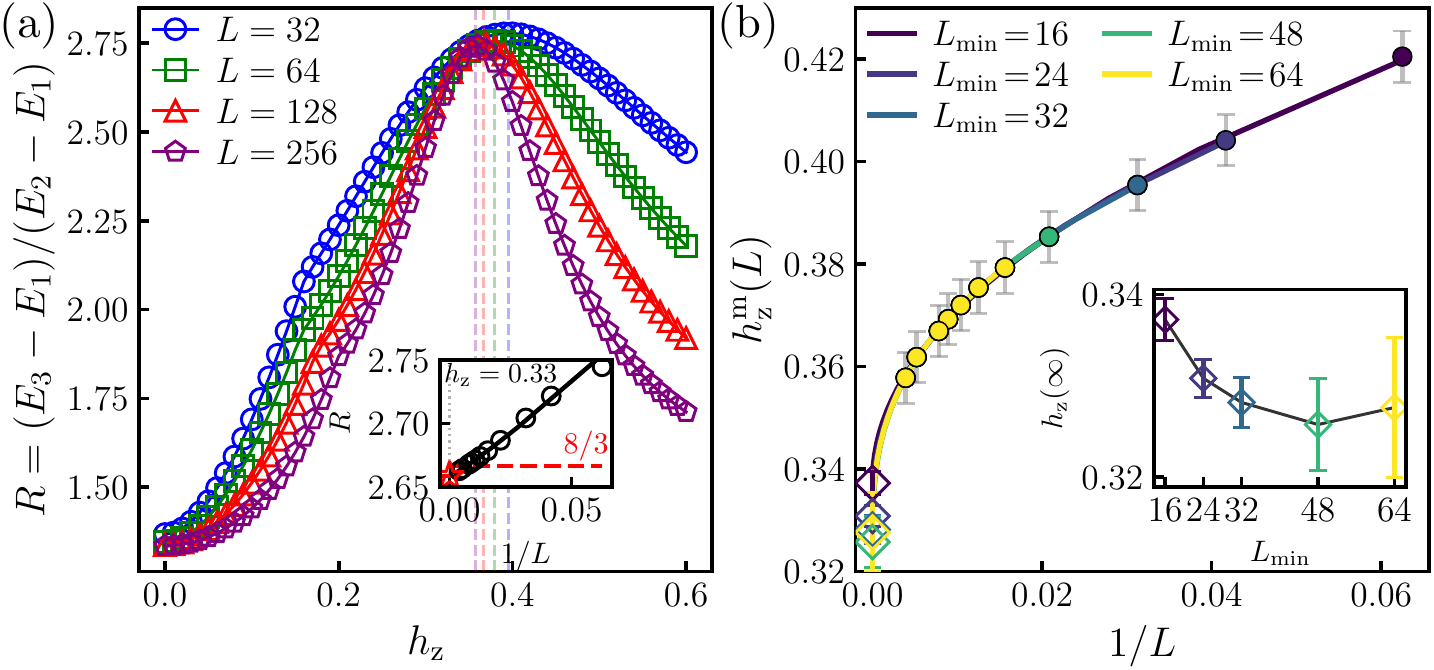}
  \caption{(a) Gap ratio $R \equiv (E_3 - E_1)/(E_2 - E_1)$ as a function of $h_\text{z}$ for the original OF model at the tricritical point, under a boundary term $-h_\text{z} (Z_1 + Z_L)$, shown for several system sizes $L$. The dashed vertical lines mark the peak positions $h_\text{z}^\text{m}(L)$, obtained by a cubic-polynomial fit around each peak. Inset: finite-size scaling of the peak ratio versus $1/L$, extrapolated via a power-law fit to $R \approx 2.66$ in the thermodynamic limit, for $h_\text{z} = 0.33$\,. (b) Extrapolation of $h_\text{z}^\text{m}(L)$ to the thermodynamic limit via a power-law fit, including data from a variable minimum size $L_\text{min}$ up to $L = 256$. The gray error bar reflects the scanning resolution of $h_\text{z}$. Inset: the extrapolated $h_\text{z}^{c}$ as a function of $L_\text{min}$, showing that the result stabilizes around $0.33$ as the smallest system sizes are successively dropped from the fit. The error bar reflects the fitting uncertainty. All simulations are performed under OBC.}
  \label{fig:free2fixed}
\end{figure}

First, we note that the free boundary condition realized by the original OF model at the tricritical point can be driven to the fixed boundary condition $|+\rangle$ by a boundary Zeeman field $-h_\text{z}(Z_{1}+Z_{L})$, which explicitly breaks the global spin-flip $\mathbb{Z}_{2}$ symmetry. 
This free-to-fixed boundary RG flow is mediated by the conformal boundary state $|0+\rangle = |\epsilon\rangle$, whose boundary operator content is given by the fusion rule $[\epsilon] \times [\epsilon] = [I] + [\epsilon']$; see Sec.~\ref{sec:tci-bcft-review} and Ref.~\cite{Iino2020PRB}. 
To determine the position of the boundary state $|0+\rangle$, we use the gap ratio $R \equiv (E_{3} - E_{1}) / (E_{2} - E_{1})$ as a diagnostic: 
the free boundary condition yields $R = 4/3$, the intermediate state $|0+\rangle$ yields $R = 8/3$, and the fixed boundary condition $|+\rangle$ yields $R = 3/2$. 
As shown in Fig.~\ref{fig:free2fixed}(a), $R$ first rises to a peak near $R \approx 8/3$ and then drops toward $R \approx 3/2$ with increasing $h_\text{z}$, signaling a boundary phase transition. 
The peak position $h_\text{z}^\text{m}(L)$ marks the finite-size estimate of the transition field. 
Furthermore, a systematic extrapolation of $h_\text{z}^\text{m}(L)$ to the thermodynamic limit [Fig.~\ref{fig:free2fixed}(b)] yields $h_\text{z}^{c} \approx 0.33$ as the transition field strength. 
Consistently, the peak ratio extrapolated to the thermodynamic limit approaches $R \approx 2.66$ at the estimated $h_\text{z}^c$, in good agreement with the prediction $R = 8/3$ for the boundary state $|0+\rangle$. 

Second, we consider a simplified cluster OF model in which the $g$ terms are truncated so as not to generate extra boundary couplings after conjugation by the open-chain SPT entangler [compared with Eq.~\eqref{eq:spt_H}]. 
The Hamiltonian reads
\begin{align}
  H_{1} = - \sum_{i=1}^{L-1} Z_{i}Z_{i+1} - \sum_{i=1}^{L-2} Z_{i}X_{i+1}Z_{i+2} + g \sum_{i=1}^{L-4} Z_{i}X_{i+1}Z_{i+3} + g \sum_{i=2}^{L-3} Z_{i}X_{i+2}Z_{i+3} - h_\text{zz} \left( Z_{1}Z_{2} + Z_{L-1}Z_{L} \right) \, ,
\label{eq:simple_H}
\end{align}
where the $g$-term summation ranges are chosen to exclude the combinations that would produce additional boundary operators. 
Conjugating with the open-chain SPT entangler $U_{\rm SPT}^{\rm OBC} = \prod_{i=1}^{L-1} \mathrm{CZ}_{i,i+1}$ yields
\begin{align}
  U_{\rm SPT}^{\rm OBC} H_{1} (U_{\rm SPT}^{\rm OBC})^\dagger = H'_\text{OF} - (h_\text{zz} + 1) \left( Z_{1}Z_{2} + Z_{L-1}Z_{L} \right) \,,
\end{align}
where $H'_\text{OF}$ is again the original OF Hamiltonian at the TCI QCP defined on sites $2$ through $L-1$. 
In contrast to the full cluster OF model, the transformed Hamiltonian contains no extra $g$-dependent boundary terms, allowing a clean analysis of the four classical sectors labeled by $z_1, z_L = \pm 1$.

\begin{figure}[t]
  \centering
  \includegraphics[width=0.6\linewidth]{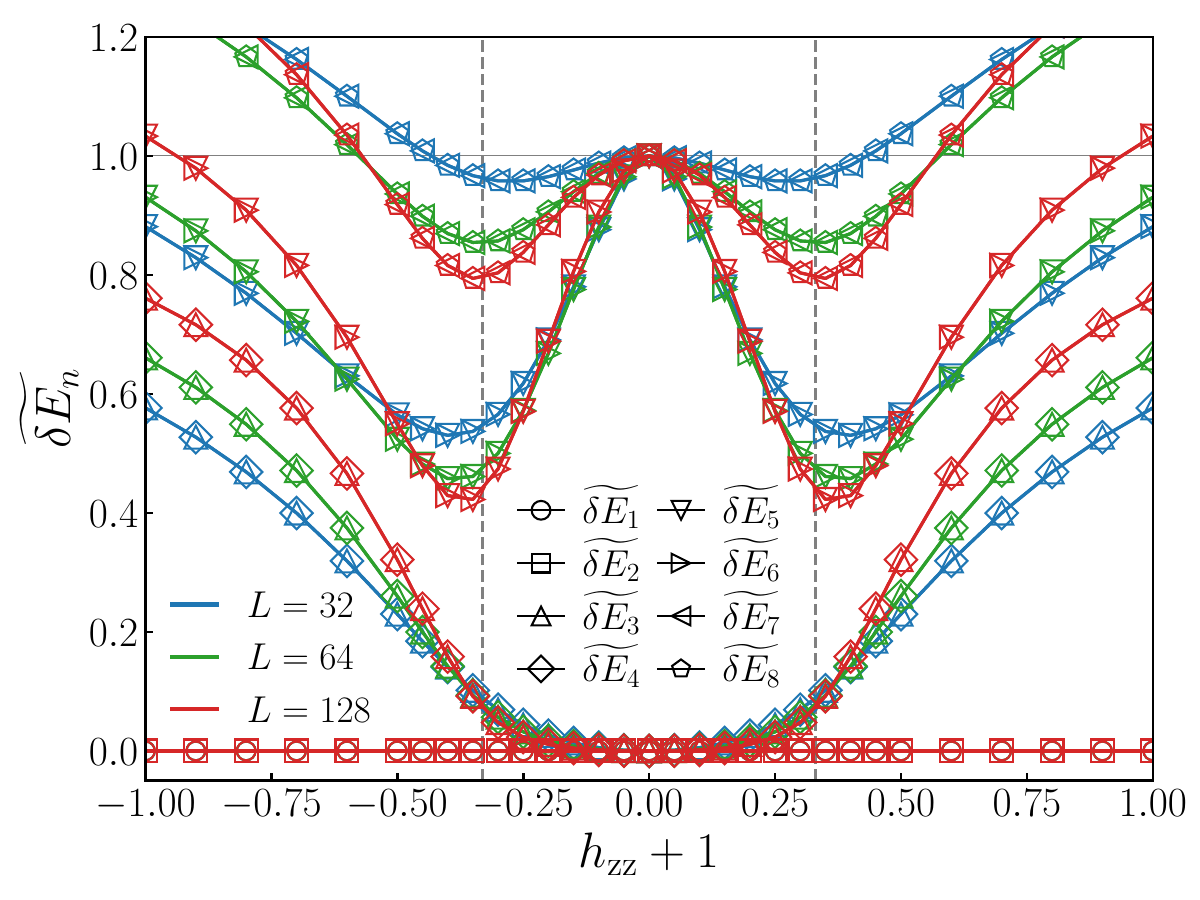}
  \caption{Low-lying energy gaps $\delta E_{n} \equiv E_{n} - E_{1}$ of the simplified cluster OF model~\eqref{eq:simple_H} as a function of $h_\text{zz}+1$, for $L = 32$, $64$, and $128$. For each system size, the whole spectrum is rescaled such that $\delta E_{5} = 1$ at $h_\text{zz} + 1 = 0$ to facilitate comparison across different sizes; $\widetilde{\delta E}_n$ denotes the rescaled gaps. Near $h_\text{zz}+1 = 0$, the four lowest energy levels become degenerate. The degeneracy is lifted when $|h_\text{zz}+1|$ exceeds a scale comparable to the transition field $h_\text{z}^{c} \approx 0.33$ determined from Fig.~\ref{fig:free2fixed} (marked by the two gray dashed lines). Simulations are performed under OBC.}
  \label{fig:fourfold_hzz}
\end{figure}

Figure~\ref{fig:fourfold_hzz} shows the resulting low-lying energy gaps as a function of $h_\text{zz}+1$. 
Near $h_\text{zz}+1 = 0$, the four lowest energy levels become degenerate; this degeneracy is lifted when $|h_\text{zz}+1|$ exceeds a scale comparable to the transition field $h_\text{z}^c \approx 0.33$ obtained from Fig.~\ref{fig:free2fixed} (marked by the gray dashed lines). 
This agreement supports the interpretation that the fourfold-degenerate window is associated with the finite basin of attraction of the free boundary condition.
As the full cluster OF model contains additional $g$-dependent boundary terms, the comparison provides numerical evidence for this boundary-RG interpretation rather than an exact identification of the transition scale.

\section{Supersymmetry under open boundary conditions}
\label{app:obc-susy}

In this appendix, we discuss the supersymmetric structure of the OF model with OBC. We first review the periodic-boundary construction and then explain the modifications required for open boundaries. 

As discussed in Refs.~\cite{OF2018PRL,zou2020prb}, the OF Hamiltonian with PBC can be written as
\begin{equation}
    H=(Q_+)^2+(Q_-)^2,\qquad 
    Q_\eta=\frac{1}{2\sqrt{\kappa}}\left[\sum_{a=1}^{N}\alpha_a\eta^a\gamma_a+\eta i\kappa\sum_{a=1}^{N}\beta_a\eta^a\gamma_{a-1}\gamma_a\gamma_{a+1}\right],
    \qquad \eta=\pm1 ,
\end{equation}
where $N=2L$ is the number of Majorana fermions, $\kappa=2g$, and the uniform OF choice is $\alpha_a=\beta_a=1$.

To evaluate $Q_+^2+Q_-^2$, we define $\Gamma_a=\gamma_{a-1}\gamma_a\gamma_{a+1}$.
Using $\{\gamma_a,\gamma_b\}=2\delta_{ab}$, one obtains
\begin{equation}
    \{\gamma_a,\Gamma_b\}=
    \begin{cases}
    2\gamma_b\gamma_{b+1}, & a=b-1,\\
    -2\gamma_{b-1}\gamma_{b+1}, & a=b,\\
    2\gamma_{b-1}\gamma_b, & a=b+1,\\
    0, & \text{otherwise},
    \end{cases}
    \qquad
    \{\Gamma_a,\Gamma_b\}=
    \begin{cases}
    -2, & a=b,\\
    2\gamma_{m-1}\gamma_m\gamma_{m+2}\gamma_{m+3}, & |a-b|=2,\quad m=\min(a,b),\\
    0, & \text{otherwise}.
    \end{cases}
\end{equation}
Writing $Q_\eta=\frac{1}{2\sqrt{\kappa}}(A_\eta+B_\eta)$, where $A$ and $B$ denote the linear and cubic parts of $Q_\eta$, we have
\begin{equation}
    Q_+^2+Q_-^2=\sum_{\eta=\pm1}\frac{1}{4\kappa}\left(A_\eta^2+\{A_\eta,B_\eta\}+B_\eta^2\right).
\end{equation}
The linear-linear part and mixed linear-cubic part give
\begin{equation}
    \sum_{\eta=\pm1}\frac{1}{8\kappa}\sum_{a,b=1}^{N}\alpha_a\alpha_b\eta^{a+b}\{\gamma_a,\gamma_b\}
    =\frac{1}{2\kappa}\sum_{a=1}^{N}\alpha_a^2,\qquad
    \sum_{\eta=\pm1}\frac{i}{4}\sum_{a,b=1}^{N}\alpha_a\beta_b\eta^{a+b+1}\{\gamma_a,\Gamma_b\}.
\end{equation}
Because of the sum over \(\eta=\pm1\), only terms with \(a+b\) odd survive. Hence the \(a=b\) term in \(\{\gamma_a,\Gamma_b\}\) drops out, and only \(a=b-1\) and \(a=b+1\) contribute. Therefore,
\begin{equation}
    \sum_{\eta=\pm1}\frac{i}{4}\sum_{a,b}\alpha_a\beta_b\eta^{a+b+1}\{\gamma_a,\Gamma_b\}
    =i\sum_{b=1}^{N}\alpha_{b-1}\beta_b\,\gamma_b\gamma_{b+1}
    +i\sum_{b=1}^{N}\alpha_{b+1}\beta_b\,\gamma_{b-1}\gamma_b
    =\sum_{j=1}^{N}t_j\,i\gamma_j\gamma_{j+1},
\end{equation}
where, for PBC, $t_j=\alpha_{j-1}\beta_j+\alpha_{j+2}\beta_{j+1}$.
The cubic-cubic part gives
\begin{equation}
\begin{aligned}
    \sum_{\eta=\pm1}\frac{-\kappa}{8}\sum_{a,b=1}^{N}\beta_a\beta_b\eta^{a+b}\{\Gamma_a,\Gamma_b\}
    &=\frac{\kappa}{2}\sum_{a=1}^{N}\beta_a^2
    -\kappa\sum_{a=1}^{N}\beta_a\beta_{a+2}\gamma_{a-1}\gamma_a\gamma_{a+2}\gamma_{a+3}.
\end{aligned}
\end{equation}
Thus
\begin{equation}
    Q_+^2+Q_-^2=C+\sum_{j=1}^{N}t_j\,i\gamma_j\gamma_{j+1}
    -\kappa\sum_{a=1}^{N}u_a\,\gamma_{a-1}\gamma_a\gamma_{a+2}\gamma_{a+3},
\end{equation}
with
\begin{equation}
    C=\frac{1}{2\kappa}\sum_{a=1}^{N}\alpha_a^2+\frac{\kappa}{2}\sum_{a=1}^{N}\beta_a^2,\qquad u_a=\beta_a\beta_{a+2}.
\end{equation}
For the uniform choice \(\alpha_a=\beta_a=1\), this reduces to
\begin{equation}
    Q_+^2+Q_-^2=\frac{N}{2\kappa}+\frac{\kappa N}{2}
    +2\sum_{j=1}^{N}i\gamma_j\gamma_{j+1}
    -\kappa\sum_{a=1}^{N}\gamma_{a-1}\gamma_a\gamma_{a+2}\gamma_{a+3}.
\end{equation}
This reproduces the periodic-boundary OF Hamiltonian, up to the additive constant and the normalization convention used in the main text.

We now consider OBC case. The natural open-chain ansatz is
\begin{equation}
    Q_\eta=\frac{1}{2\sqrt{\kappa}}\left[\sum_{a=1}^{N}\alpha_a\eta^a\gamma_a+\eta i\kappa\sum_{a=2}^{N-1}\beta_a\eta^a\gamma_{a-1}\gamma_a\gamma_{a+1}\right],
    \qquad \eta=\pm1 .
\end{equation}
Repeating the same derivation, but without periodic wrap-around terms, gives
\begin{equation}
    Q_+^2+Q_-^2=C_{\rm OBC}+\sum_{j=1}^{N-1}t_j\,i\gamma_j\gamma_{j+1}
    -\kappa\sum_{a=2}^{N-3}u_a\,\gamma_{a-1}\gamma_a\gamma_{a+2}\gamma_{a+3},
\end{equation}
where
\begin{equation}
    C_{\rm OBC}=\frac{1}{2\kappa}\sum_{a=1}^{N}\alpha_a^2+\frac{\kappa}{2}\sum_{a=2}^{N-1}\beta_a^2,\qquad u_a=\beta_a\beta_{a+2},\qquad
    t_j=\mathbf 1_{2\le j\le N-1}\alpha_{j-1}\beta_j+\mathbf 1_{1\le j\le N-2}\alpha_{j+2}\beta_{j+1}.
\end{equation}
Equivalently, the hopping term coefficients are
\begin{equation}
    t_1=\alpha_3\beta_2,\qquad
    t_j=\alpha_{j-1}\beta_j+\alpha_{j+2}\beta_{j+1}\quad(2\le j\le N-2),\qquad
    t_{N-1}=\alpha_{N-2}\beta_{N-1}.
\end{equation}

Using the Jordan--Wigner convention $i\gamma_{2j-1}\gamma_{2j}=-X_j$ and $i\gamma_{2j}\gamma_{2j+1}=-Z_jZ_{j+1}$, we obtain
\begin{equation}
\begin{aligned}
    Q_+^2+Q_-^2
    =C_{\rm OBC}-\sum_{j=1}^{L}t_{2j-1}X_j-\sum_{j=1}^{L-1}t_{2j}Z_jZ_{j+1}+\kappa\sum_{j=1}^{L-2}u_{2j}X_jZ_{j+1}Z_{j+2}
    +\kappa\sum_{j=1}^{L-2}u_{2j+1}Z_jZ_{j+1}X_{j+2}.
\end{aligned}
\end{equation}
For the uniform open-chain choice \(\alpha_a=\beta_a=1\), one finds $u_a=1,\ t_1=t_{N-1}=1,\ t_j=2\quad(2\le j\le N-2)$.
Therefore,
\begin{equation}
\begin{split}
    Q_+^2+Q_-^2=&C_{\rm OBC}+X_1+X_L-2\sum_{j=1}^{L}X_j-2\sum_{j=1}^{L-1}Z_jZ_{j+1}+\kappa \sum_{j=1}^{L-2}\left(X_jZ_{j+1}Z_{j+2}+Z_jZ_{j+1}X_{j+2}\right)\\
    =&C_{\rm OBC}+H_{\rm std}^{\rm OBC}+X_1+X_L,
\end{split}
\end{equation}
where $C_{\rm OBC}=\frac{N}{2\kappa}+\frac{\kappa(N-2)}{2}$, and $H_{\rm std}^{\rm OBC}$ is the truncated open-chain OF model Hamiltonian.
Thus the truncated open-chain supercharge gives the same bulk Hamiltonian as the OF model, but add a boundary transverse fields with coefficient $1$.

The boundary $ZZ$ couplings can also be tuned by changing $\alpha_1$ and $\alpha_N$.
Keeping $\beta_a=1$ and $\alpha_a=1$ for $a\neq 1,N$, we get $t_1=t_{N-1}=1,\ t_2=1+\alpha_1,\ t_{N-2}=1+\alpha_N$, while all bulk coefficients remain unchanged. 
Therefore,
\begin{equation}
    Q_+^2+Q_-^2=C_{\rm OBC}'+H_{\rm std}^{\rm OBC}+X_1+X_L-(\alpha_1-1)Z_1Z_2-(\alpha_N-1)Z_{L-1}Z_L,
\end{equation}
where $C_{\rm OBC}'=\frac{N-2+\alpha_1^2+\alpha_N^2}{2\kappa}+\frac{\kappa(N-2)}{2}$.

Finally, the identity \(H=(Q_+)^2+(Q_-)^2\) does not uniquely determine \(Q_\eta\). For example, if \(\beta_a=1\) is fixed, there are \(N\) coefficients \(\alpha_a\), while the Hamiltonian only fixes the \(N-1\) bilinear coefficients \(t_j\). Thus a one-parameter family of \(\alpha_a\)'s can give the same Hamiltonian terms, up to a different additive constant. With general \(\beta_a\), this redundancy is even larger. However, not every algebraic square root should be identified with the physical lattice supersymmetry current. A candidate \(Q_\eta\) represents the lattice supercurrent only if its local density flows to the continuum supercurrent in the low-energy limit. Therefore, finding the physically appropriate open-chain lattice supercharge requires more than solving \(H=(Q_+)^2+(Q_-)^2\); it also requires checking continuum supercurrent matrix elements, or equivalently the corresponding super-Virasoro structure.

\bibliography{main}

\end{document}